\documentclass[preprint,10pt]{elsarticle}
\usepackage{geometry}
\usepackage{graphicx,amsfonts,amsmath,bm,epsfig,color,latexsym}
\usepackage{epstopdf}
\usepackage{subfig}
\usepackage{float}
\usepackage{float}
\usepackage{natbib}
\usepackage[left]{lineno}
\renewcommand{\b}{\beta}
\newcommand{\om}{\omega}
\begin{document}
	
	\begin{frontmatter}
		
		\title{Dipole and quadrupole nonparaxial solitary waves}
		
		\author[a]{Naresh Saha}
		\ead{naresh\_r@isical.ac.in}
		
		\author[a]{Barnana Roy\corref{cor1}}
		\ead{barnana@isical.ac.in}
		\address[a]{Physics and Applied Mathematics Unit, Indian Statistical Institute, Kolkata-700108, India.}
		\author[b]{Avinash Khare}
		\ead{avinashkhare45@gmail.com}
		\address[b]{Department of Physics, Savitribai Phule Pune University, Pune - 411007, India.}
		\cortext[cor1]{Corresponding author}


\begin{abstract}
The cubic nonlinear Helmholtz equation with third and fourth order dispersion and non-Kerr nonlinearity like the self steepening and the self frequency shift is considered. This model describes nonparaxial ultrashort pulse propagation in an optical medium in the presence of spatial dispersion originating from the failure of slowly varying envelope approximation. We show that this system admits periodic (elliptic) solitary waves with dipole structure within a period and also transition from dipole to quadrupole structure within a period depending on the value of the modulus parameter of Jacobi elliptic function. The parametric conditions to be satisfied for the existence of these solutions are given. The effect of the nonparaxial parameter on physical quantities like amplitude, pulse-width and speed of the solitary waves are examined. It is found that by adjusting the nonparaxial parameter, the speed of solitary waves can be decelerated. The stability and robustness of the solitary waves are discussed numerically.
\end{abstract}

\end{frontmatter}

\makeatletter
\def\ps@pprintTitle{%
\let\@oddhead\@empty
\let\@evenhead\@empty
\def\@oddfoot{\reset@font\hfil\thepage\hfil}
\let\@evenfoot\@oddfoot
}
The propagation of broad self-trapped beams (pulses) in Kerr-type nonlinear media in the presence of spatial dispersion arising from the nonparaxial effect, is described by the nonlinear Helmholtz equation. Such an effect is important in the case when the wavelength is comparable to the beam (pulse) width, e.g. in the progressive miniaturization of the optical devices. For studying the propagation of ultrashort  nonparaxial  optical beams (pulses), the higher order dispersion/diffraction terms as well as non-Kerr nonlinear terms like the self steepening and the self frequency shift should be included in the nonlinear Helmholtz equation. We present analytical periodic (elliptic) solitary waves which exhibit the dipole and the quadrupole structure within a period. Importantly, transition from the dipole to the quadrupole structure is possible depending on the value of the modulus parameter of Jacobi elliptic function. Numerical simulations validate the stability and robustness of the solitary wave solutions. 

\label{}
	
   \section {Introduction:}\label{in}
	In nonlinear optics, the study of solitons has been a subject of intense investigations since their first theoretical prediction \cite{chiao}. It is widely accepted that the cubic nonlinear Schrodinger equation (NLSE) is the generic model for the description of dynamics of optical solitons in Kerr type nonlinear media \cite{hase4}.
	Hasegawa and Tappert derived the NLSE by employing paraxial approximation or slowly varying envelope approximations (SVEA) \cite{has}, in which the width of a sufficiently low intensity pulse (beam) is larger than the carrier wavelength and it propagates along (or at negligible angles) a reference axis. If at least one of the conditions is not met, the pulse (beam) is referred to as nonparaxial \cite{cham11}. Interest on nonparaxial beams has started attracting interest after the pioneering work \cite{lax} in which the propagation of ultra narrow optical beams is considered. On the other hand, Helmholtz nonparaxiality occurs when the broad pulse (beam) propagates at arbitrary angles to the reference direction \cite{cham45,cham11,cham46}. The governing equation is then known as the Helmholtz-Manakov (HM) equation \cite{chris06}. In both of the above scenarios, the equivalence of the transverse and longitudinal dimensions in uniform media is achieved by ignoring SVEA.\\ 
	The use of ultra narrow soliton beams in miniaturized optical devices is proposed in \cite{cham45}. 
	The analytical soliton solutions to the scalar NLH equation are derived in \cite{cham11,cham19} for focusing and defocusing Kerr nonlinearity, while those for the power law, polynomial, and the saturable nonlinearity are obtained in \cite{christ77,christ20,chris09}, those in the anomalous and normal group velocity dispersion regimes are obtained in \cite{chris71}. 
	Further, exact soliton solutions are studied in the relativistic and pseudo relativistic formulation of the scalar NLH equaion with cubic, quintic and saturable nonlinearity \cite{chris66,mc2012}. Quasi soliton behavior of the exact as well as approximate solutions to the scalar and vector nonparaxial evolution equations developed for describing propagation in two dimensions, are shown in \cite{blair}. 
Vortex solitons in two dimensions \cite{wang}, rogue waves \cite{este}, and numerical solutions to the nonparaxial nonlinear Sch\"odinger equation are obtained in \cite{mala}. 
	 The study of nonparaxiality in $\cal{PT}$-symmetric optical context is found in \cite{karta}. Solitary wave solutions to the coupled NLH equation in $(1+1)$ dimensions and to the coupled nonparaxial NLSE in $(2+1)$ dimensions are obtained in \cite{kanna} and \cite{kumar} respectively. The effect of nonparaxiality on the amplitude, pulse-width and the speed of the solitary waves are studied too \cite{kanna}. 
	 Modulational instability of the cubic quintic NLH system is studied in \cite{tamil} and the chirped elliptic and hyperbolic solitary wave solutions are presented while those for the cubic coupled NLH equation in the presence of the self steepening (SS) and the self frequency (SFS) shift are studied in \cite{naresh}. 
	
	However, for modeling the propagation of sub-picosecond (femtosecond) optical pulses (beams) in non-paraxial regime, it is required to incorporate higher order effects, such as the third order dispersion (TOD), the fourth order dispersion (FOD) and non-Kerr nonlinear terms like the SS and the SFS into the scalar NLH equation. 
	It is relevant to mention that different higher order effects appear in optical materials for various applications such as, in ultrahigh-bit-rate optical communication systems, ultrafast physical processes etc.\cite{krug72}. TOD plays an important role in the propagation of short pulses whose widths are nearly $50$ fs \cite{pala1,pala2}. When pulse widths are shorter than $10$ fs, FOD becomes essential \cite{pala1,pala2}. The wave dynamics in such a situation in paraxial regime, is governed by higher order NLSE  (HNLSE) \cite{koda5}. Recent researches have focused on seeking more complex, such as,  dipole, multipole structures \cite{he23} than nodeless amplitude distributions featured by fundamental solitons for HNLSE. It is relevant to mention that a bright solitary wave (having a $sech$-profile) has a localized intensity peak over a continuous wave background and the intensity profile vanishes at $\pm \infty$ whereas a dark solitary wave (having $\tanh$-profile) exhibits a dip in a continuous wave background and the intensity profile increases monotonically from the dip at the center and approaches a constant value at $\pm \infty$. When the intensity profile of a solitary wave has two peaks and one dip at the centre (where the light intensity is zero) in a continuous wave background, it is referred to as dipole solitary wave. It can be viewed as an inter-modulation of bright and dark components with the same pulse width and velocity \cite{mod2016}.  When the intensity profile of a solitary wave has four peaks and three dips in a continuous wave background, it is referred to as quadrupole solitary wave. As mentioned in \cite{ol2004,ye,chou} multipole solitary waves can be viewed as composed of several fundamental solitary waves with alternating phases. This opens up the possibility of encoding complex solitary images and letters by packing an arbitrary number of fundamental solitary waves into a stable matrix \cite{ol2004}.
	 Dipole solitons occur in many optical media, like, nonlocal nonlinear media \cite{ge,jung,huang,shen}, thermal nonlinear media \cite{ye} and photonic lattices \cite{sus,xu1}. Stationary two dimensional multipole mode solitons have been observed experimentally in a medium with thermal nonlinearity \cite{rot}.
	Efficient ansatz approaches for example, sum and product of bright and dark solitons have been used to derive the
	so-called dark-in-the -bright, dipole, tripole, and the multipole solutions \cite{li,yang,hong,azz,chou,triki,azz1,tian}. Absence of such type of study in the nonparxial regime, so far as we are aware of, has motivated us to look for dipole and multipole solitary waves in NLH system with higher order dispersion and non-Kerr nonlinearity. As it extends the known classification of solitary wave solution namely dipole and multipole solitary wave to the nonparaxial regime, this is a problem well worth investigating.\\
	The objective of the present study is to investigate the solitary wave propagation for higher order NLH systems with cubic Kerr nonlinearity and spatial dispersion which arises when SVEA fails.
    By considering the effect of TOD, FOD, SS and SFS, we show that the considered system admits dipole soliton and periodic solitary waves with the dipole and quadrupole structure within a period under certain parametric conditions. The novelty of the present study lies not only in the existence of dipole structure within a period of periodic (elliptic) solitary wave but also in the transition from dipole to quadrupole structure within a period, depending on the value of the modulus parameter $m$ of the Jacobi elliptic function \cite{abra}. So far as we are aware of, these interesting propagation characteristics of the nonparaxial periodic waves is uncovered for the first time in this article. The evolution and stability of these waves are analyzed by direct numerical simulations.\\
	 The organization of the present article is as follows. In Section II we present the theoretical model of the cubic nonlinear Helmholtz equation in the presence of TOD, FOD, SS and SFS. This model describes propagation of ultrashort nonparaxial solitary waves in optical waveguide/fibre with the spatial dispersion arising from the failure of SVEA. In Section III, we obtain the dipole and quadrupole solitary waves. The characteristic features of these solitary waves and their physical significance are studied too. In Section IV we discuss the stability and robustness of the four solutions obtained here. Finally, in Section V, we summarize our results and indicate their relevance.

	\section{Theoretical model}\label{model}
	We start with the following higher order dimensionless NLH equation describing pulse propagation in the nonparaxial regime \cite{mc2012}
	\begin{equation}
		\begin{array}{lcl}\label{1.0}
			i(\frac{\partial \psi}{\partial \zeta} + \alpha \frac{\partial \psi}{\partial \tau}) + \frac{a_1}{2} \frac{\partial^2 \psi}{\partial \tau^2} + a_2 |\psi|^2 \psi + i[a_3 \frac{\partial^3 \psi}{\partial \tau^3}+ a_4 \frac{\partial (|\psi|^2 \psi)}{\partial \tau} + a_5 \psi \frac{\partial (|\psi|^2)}{\partial \tau}] + a_6 \frac{\partial^2 \psi}{\partial \zeta^2} + a_7 \frac{\partial^4 \psi}{\partial \tau^4} = 0
		\end{array}
	\end{equation}
where $\psi(\zeta,\tau)$ is the complex pulse envelope, $\zeta$ and $\tau$ are dimensionless space and time variables respectively. The coefficients $a_i, i=1,2,3,7$ are respectively the group velocity dispersion (GVD) , the self phase modulation (SPM), the TOD, and the FOD. $a_2 = 1$, $a_1 = \pm 1$ indicates anomalous and normal GVD respectively. The term proportional to $a_4$ corresponds to 
SS \cite{anda}. This causes a temporal pulse distortion, which leads to the formation of an optical shock at the trailing edge of the pulse \cite{oliv} unless balanced by the dispersion. The term proportional to $a_5$ accounts for the delayed Raman response \cite{yui} which is responsible for the pulse undergoing a frequency shift known as SFS \cite{mit}. The term proportional to $a_6 (>0)$ (nonparaxial parameter) refers to the spatial dispersion arising from the nonparaxial effect.\\
The term $i\alpha \frac{\partial \psi}{\partial \tau}$ in (\ref{1.0}) can be transformed away by the Galilean boost
\begin{equation}
t = \tau - \alpha \zeta, ~~~~ x = \zeta
\end{equation}
i.e. the solutions to (\ref{1.0}) describe waves in a frame moving with speed $\frac{1}{\alpha}$ along the $\zeta$ axis. Thus Eq. (\ref{1.0}) is transformed into 
\begin{equation}\label{1.1a}
	\begin{split}
	i\frac{\partial \psi}{\partial x} + \frac{1}{2}(a_1 + 2 a_6 \alpha^2) \frac{\partial^2 \psi}{\partial t^2} - 2 \alpha a_6 \frac{\partial^2 \psi}{\partial x \partial t} +  a_2 |\psi|^2 \psi + i[a_3 \frac{\partial^3 \psi}{\partial t^3}+ a_4 \frac{\partial (|\psi|^2 \psi)}{\partial t} + a_5 \psi \frac{\partial (|\psi|^2)}{\partial t}]\\ + a_6 \frac{\partial^2 \psi}{\partial x^2} + a_7 \frac{\partial^4 \psi}{\partial t^4} = 0
	\end{split}
\end{equation}
Now we restrict to solutions for which $\alpha a_6 << O(1)$ and $a_6 \alpha^2 << O(1)$. Also the terms $\frac{\partial^2 \psi}{\partial t^2}$ and $\frac{\partial^2 \psi}{\partial x \partial t}$ are assumed to be both $O(1)$. These solutions will satisfy the equation
\begin{equation}\label{1.1b}
	\begin{array}{lcl}
		i\frac{\partial \psi}{\partial x} + \frac{a_1}{2} \frac{\partial^2 \psi}{\partial t^2} +  a_2 |\psi|^2 \psi + i[a_3 \frac{\partial^3 \psi}{\partial t^3}+ a_4 \frac{\partial (|\psi|^2 \psi)}{\partial t} + a_5 \psi \frac{\partial (|\psi|^2)}{\partial t}] + a_6 \frac{\partial^2 \psi}{\partial x^2} + a_7 \frac{\partial^4 \psi}{\partial t^4} = 0
	\end{array}
\end{equation}

	 In what follows we shall take $a_1, a_2 = 1$. It is important to note that the term $a_6 \frac{\partial^2 \psi}{\partial x^2}$ in (\ref{1.1b}) arising from the absence of SVEA, leads to a dispersion relation that supports both forward and backward propagating waves. This is in sharp contrast to NLS \cite{zak1973} and Manakov \cite{mana1974} equations where the backward propagating waves are not allowed \cite{chris06}.
	
	To obtain the dipole solitary waves we use the following traveling wave ansatz
	\begin{equation}\label{1.2}
		\psi = u(\xi)e^{i\phi(x,t)},~~\phi(x,t)=k x-\omega t+\phi_0,~~\xi = \beta(t-cx),	
	\end{equation}
	where $u$ is the amplitude and $\phi$ is the phase of the solitary wave. These are real functions of $\xi$. The parameter $c, k, \phi_0$ are the inverse velocity, wave number and real constant respectively. 
	On substituting this ansatz in Eq. (\ref{1.1b}) and separating the real and imaginary parts we obtain
	\begin{equation}\label{1.5}
		\begin{array}{lcl}
			a_7 \beta^4 u''''+ (\frac{1}{2}a_1 + a_6 c^2 + 3 a_3 \omega -6 a_7 \omega^2) \beta^2 u''+(a_2 + 
			\omega a_4)u^3 -(k + k^2 a_6 + \frac{1}{2} \omega^2a_1 + \omega^3 a_3 - \omega^4 a_7) u = 0\,,
		\end{array}
	\end{equation} 
	and further
	\begin{equation}\label{1.6}
		\begin{array}{lcl}
			(a_3 -4\omega a_7) \beta^3 u''' +(3a_4 + 2a_5)\beta u^2 u' 	-(c+2 k c a_6 + \omega a_1 +3\omega^2 a_3 -4\omega^3 a_7) \beta u' = 0\,
		\end{array}
	\end{equation}
	\section{Dipole/quadrupole solitary waves}\label{sol}
	{\bf Solution I. Dipole Soliton}. It is easily checked by direct substitution that Eqs. (\ref{1.5}) and (\ref{1.6}) admit the following dipole solitary wave solution\\
	\begin{equation}\label{1.12}
		u(\xi) = A sech(\xi) \tanh(\xi)\,,
	\end{equation}
	provided
	\begin{subequations}\label{6}
		\begin{align}
			2a_5+3a_4 & = 0 \\ 
			a_3-4\om a_7 & = 0 \\ 
			(1+2ka_6)c & = -\om(a_1+8\om^2a_7) \\ \
			(a_2+\om a_4)A^2 & = 120 a_7\b^4 
		\end{align}
	\end{subequations}
	and further
	\begin{equation}\label{8}
		\begin{array}{lcl}
			\frac{1}{2}a_1+c^2a_6 =-2(3\om^2-5\b^2)a_7\\
			k+k^2a_6+\frac{1}{2}\om^2a_1 =(11\b^4-3\om^4)a_7
		\end{array}
	\end{equation}
It is relevant to mention that similar constraint relation between the SS and the SFS (Eqn.\ref{6}(a)) has also been obtained in \cite{li,Rel2,chou,triki,Rel5} for obtaining different solitary wave solutions of nonlinear Sch\"odinger equation containing higher order dispersion and non-Kerr nonlinearities and also for obtaining chirped solitary wave solutions of coupled cubic NLH equation in the presence of SS and SFS \cite{naresh}.\\
	We now show that the dipole solitary wave 
	(\ref{1.12}) is indeed a dipole soliton by considering the collision 
	dynamics as depicted in Fig. 1.
	For this the initial condition chosen is
	\begin{equation}\label{12}
		\begin{array}{rcr}
			\Psi(0,t) = \psi(0,t) -1.5~sech(0.2t+5)~\tanh(0.2t+5) \displaystyle e^{i(-\om t+\phi_0)}
		\end{array}
	\end{equation}
	The two dipole solitary waves converge during propagation and at a distance of about $x = 10~ \rm units$ they collide as shown in Fig.1. It is quite evident that the two dipole solitary waves collide elastically and continue their propagation after the collision. The intensity profile of the dipole soliton (\ref{1.12}) with respect to $t$ is shown in Fig.2(a) while the numerical simulation of the intensity profile is shown in Fig.2(c) demonstrating its stable evolution for the chosen values of the parameters. From Fig.2(b) we can see that as the nonparaxial parameter $a_6$ increases, the amplitude $A$ increases but the speed $|c|$ and the pulse width $\frac{1}{\beta}$ both decrease.
	
	\begin{figure}[]
		\centering
		\includegraphics[width=6cm,height=3cm]{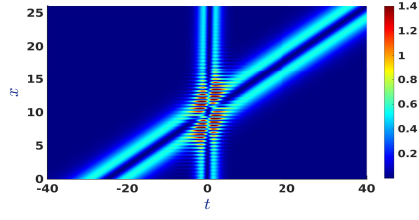}
		\caption{Evolution of the intensity profile and collision dynamics of (\ref{1.12}) as a function of propagation distance $x$ for $a_3=\frac{1}{60},~a_4=0.126,~a_5=-0.189,~a_6=0.5,~a_7=0.15,\beta=0.577914,~\omega=0.027778,~\phi_0=0.5,~c=0.023777,~k=-2.169328,~A=1.4195$.
		}
	\end{figure}

	\begin{figure}
			\centering
			\subfloat[]{\includegraphics[width=4.5cm,height=4.0cm]{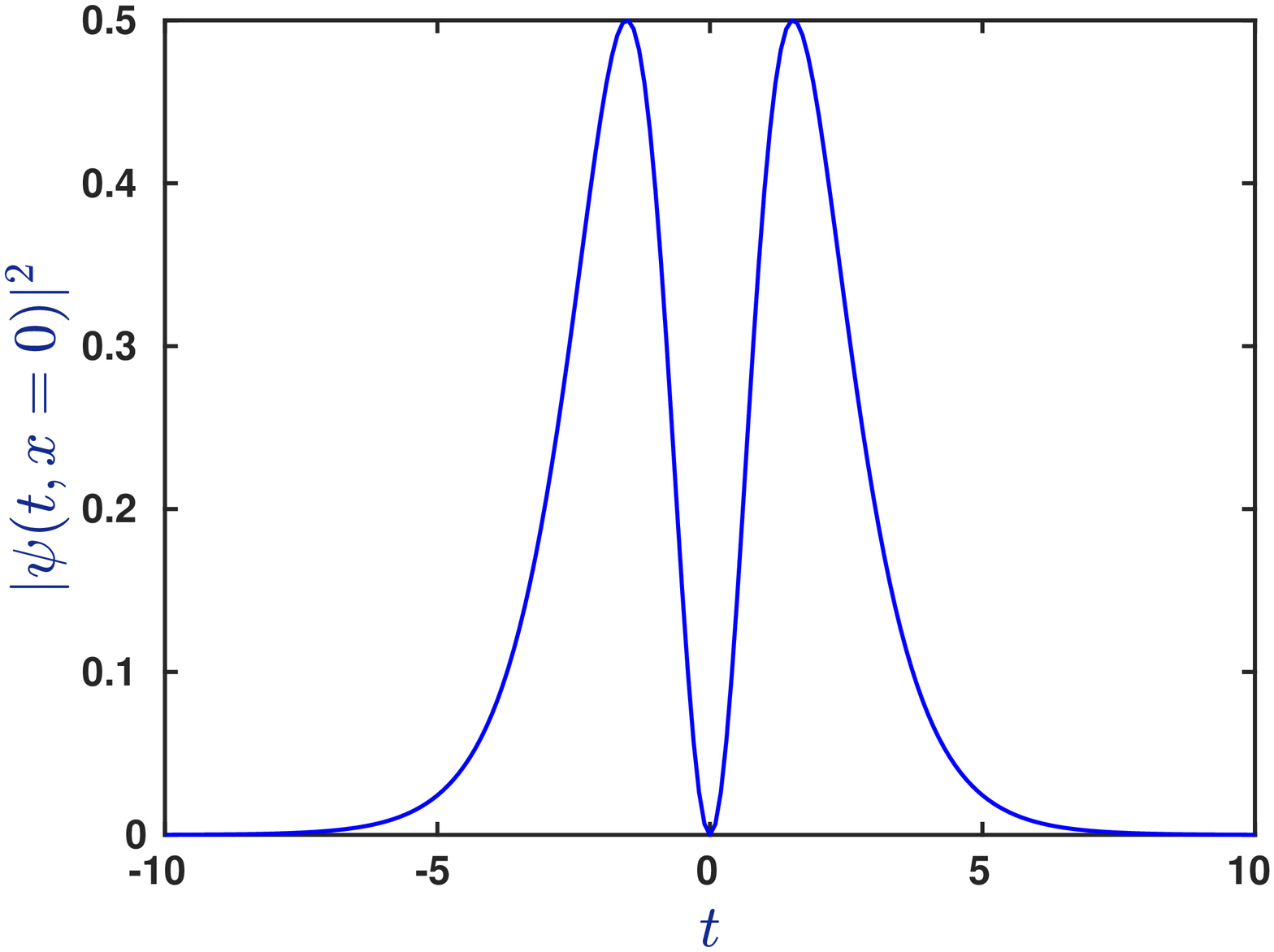}}
			~~
			\subfloat[]{\includegraphics[width=4.5cm,height=4.0cm]{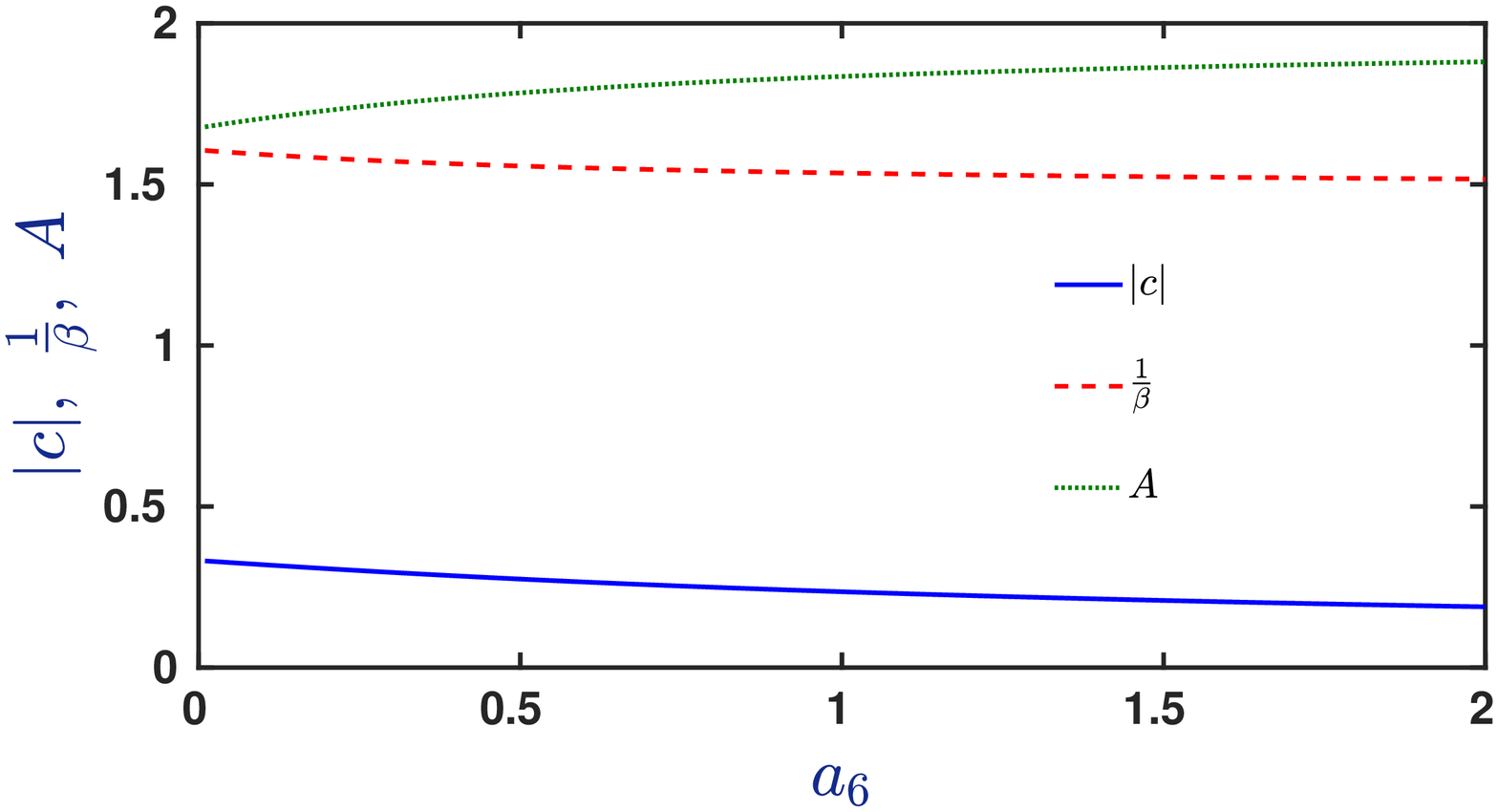}}
			~~
			\subfloat[\label{}]{\includegraphics[width=4.5cm,height=4.0cm]{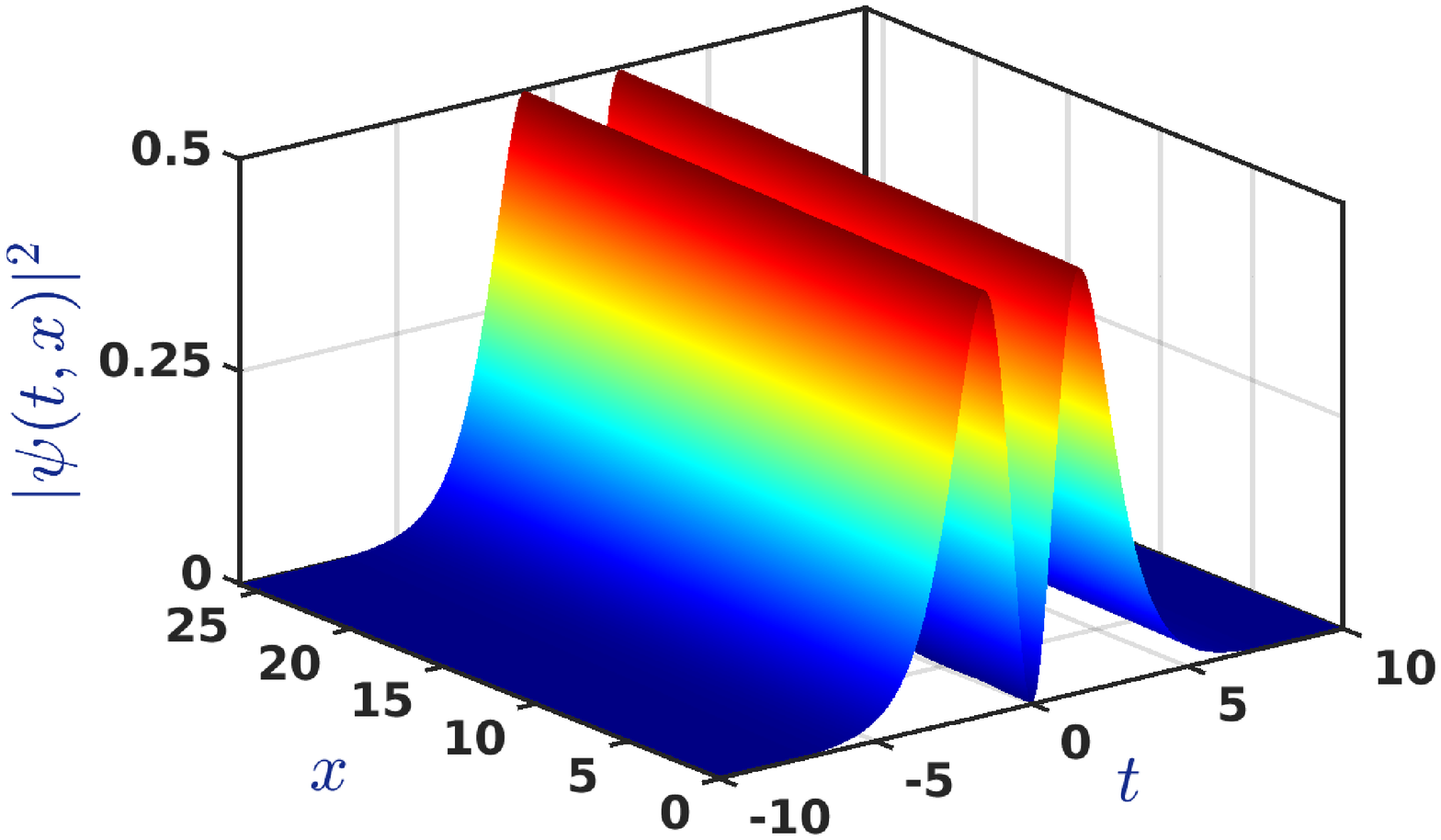}}\\
			~~
			\subfloat[\label{}]{\includegraphics[width=4.5cm,height=4.0cm]{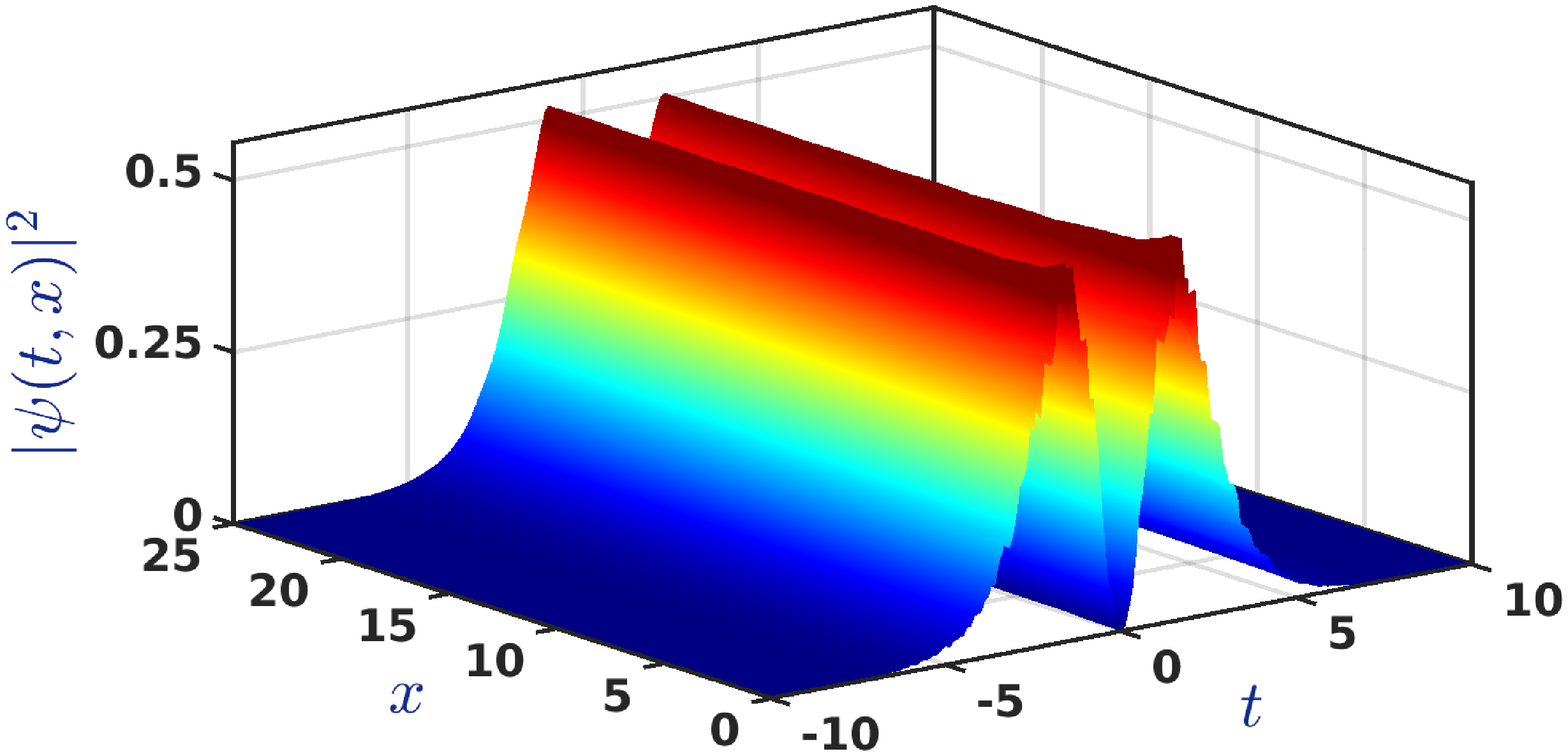}}
			~~
			\subfloat[\label{}]{\includegraphics[width=4.5cm,height=4.0cm]{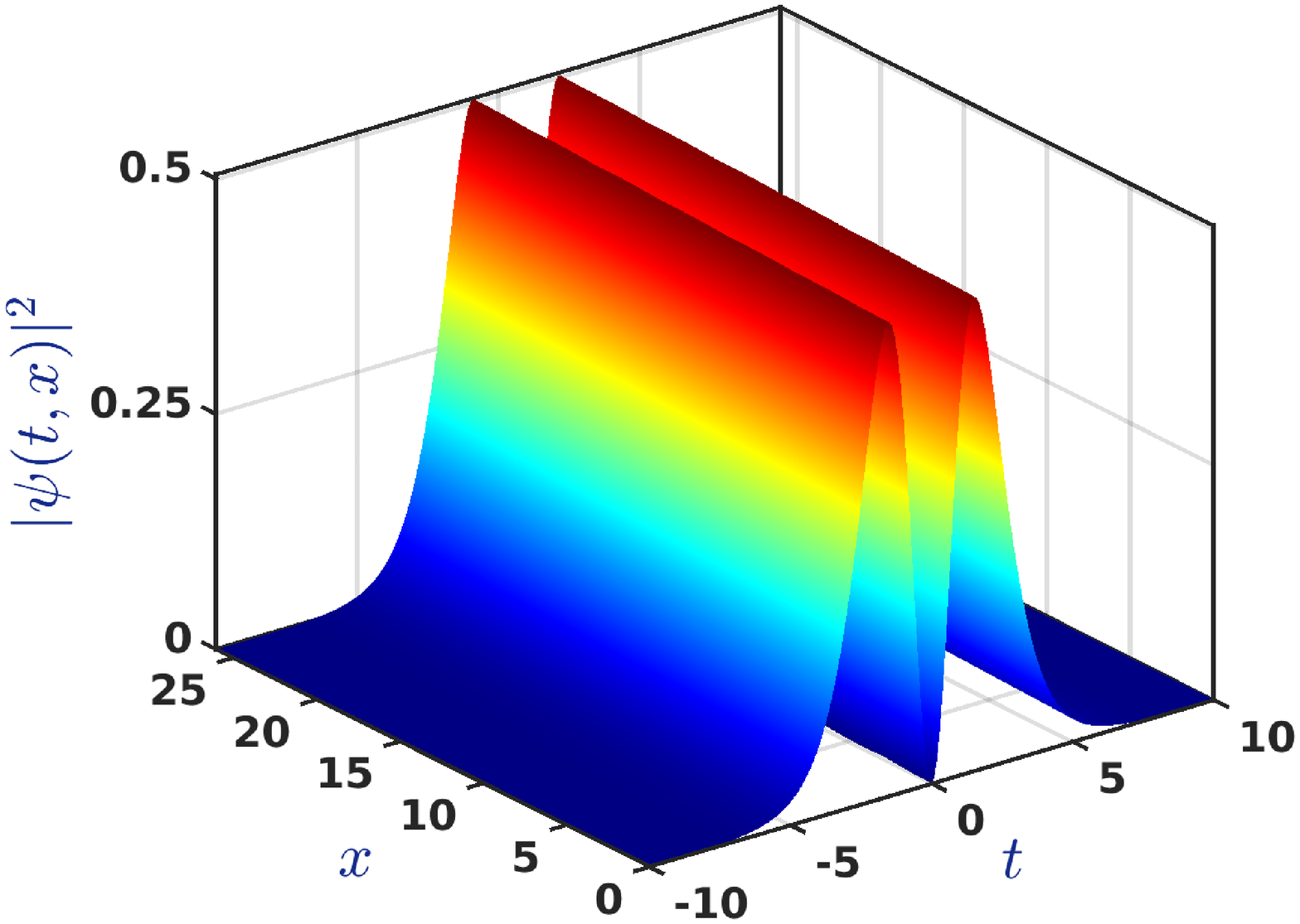}}
			~~
			\subfloat[\label{}]{\includegraphics[width=4.5cm,height=4.0cm]{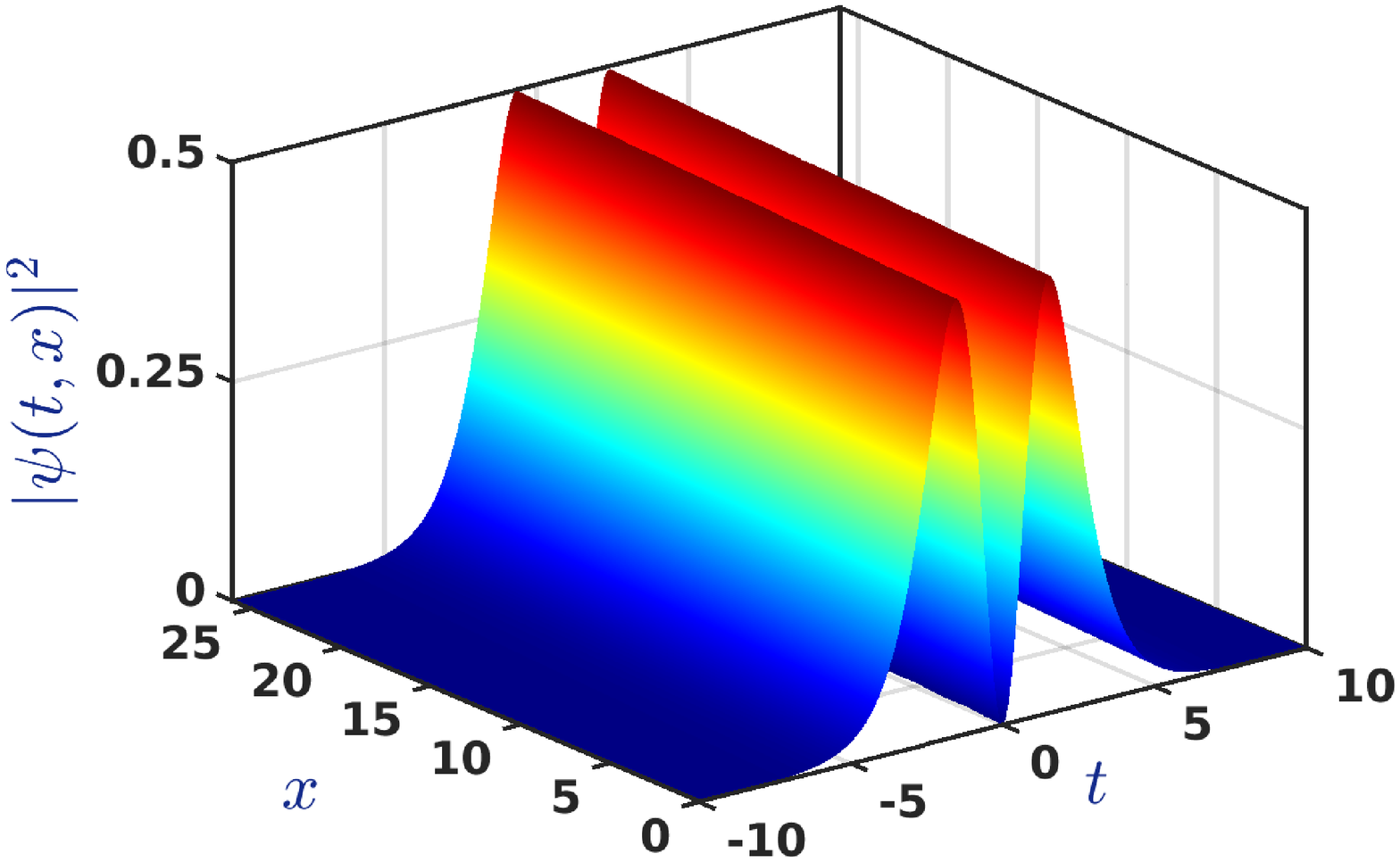}}
			\caption{(a) plot of $|\psi|^2$ versus $t$ at $x=0$ (c), (d) numerical simulations of the intensity profile without and with $10\%$ white noise respectively of the solution (\ref{1.12}) for $a_3=\frac{1}{60},~a_4=0.126,~a_5=-0.189,~a_6=0.5,~a_7=0.15,\beta=0.577914,~\omega=0.027778,~\phi_0=0.5,~c=0.023777,~k=-2.169328,~A=1.4195$ respectively, (b) plots of $|c|$, $\frac{1}{\beta}$, $A$ versus $a_6$ for the solution (\ref{1.12}) for $a_3=-0.18,~a_4=0.126,~a_7=0.15$, (e) plot of the numerical simulation of the intensity profile by adding $0.003$ to the parameters $a_i, i = 1,2, \cdots 7$ and (f) plot of the numerical simulation of the intensity profile by adding the overall factor $0.001$ to the parametric conditions (\ref{6}) \& (\ref{8}) respectively.}
	\end{figure}

	{\bf Solution II. Periodic wave exhibiting dipole/quadrupole structure within a period}\\ \\
	Eqns.(\ref{1.5}) and (\ref{1.6}) also admit the following elliptic solitary wave solution 
	\begin{equation}\label{1.20}
		u(\xi) = A\sqrt{m}  sn(\xi, m) dn(\xi, m)\,,
	\end{equation}
	 ($m$ being the modulus of the Jacobi elliptic functions $sn, dn$, $0< m <1$ \cite{abra}) provided Eqs.(6a) - (6d) are satisfied and in addition, the following two relations are also satisfied 
	\begin{equation}\label{1.21}
		\begin{array}{lcl}
			\frac{1}{2}a_1 + c^2 a_6 = -2[3\omega^2-5(2m-1) \beta^2] a_7\\
			k + k^2 a_6 + \frac{1}{2}\omega^2a_1 = [(11+64m -64 m^2) \beta^4 - 3\omega^4] a_7
		\end{array}
	\end{equation}
	It is easily checked that this solution exhibits dipole structure within a period so long $m\leq 0.5$. On the other hand, for values of $m > 0.5$, it exhibits quadrupole structure within a period. Fig.3(a) and 4(a) respectively depict the dipole and the quadrupole intensity profiles of (\ref{1.20}) respectively while Figs. 3(c) and 4(c) depict the numerical simulation of the intensity profiles of the dipole and quadrupole structure respectively. As is evident from the figures, both the dipole and the quadrupole structures exhibit stable evolutions for the chosen parameters. From Figs.3(b) and 4(b), we can see that the speed $|c|$ of both the elliptic waves having dipole and quadrupole structure within a period decreases as $a_6$ increases. Fig.3(b) shows that both the amplitude and the pulse width of the elliptic wave having dipole structure within a period increase for increasing $a_6$ but for elliptic wave having quadrupole structure within a period, while the amplitude increases, the pulse width decreases as $a_6$ increases as shown in Fig. 4(b) . It is worth noting that the solution (\ref{1.12}) is the $m=1$ limit of the elliptic wave solution (\ref{1.20}).\\ \\
	
	\begin{figure}[]
		\begin{center}
			\subfloat[\label{}]{\includegraphics[width=4.5cm,height=4.0cm]{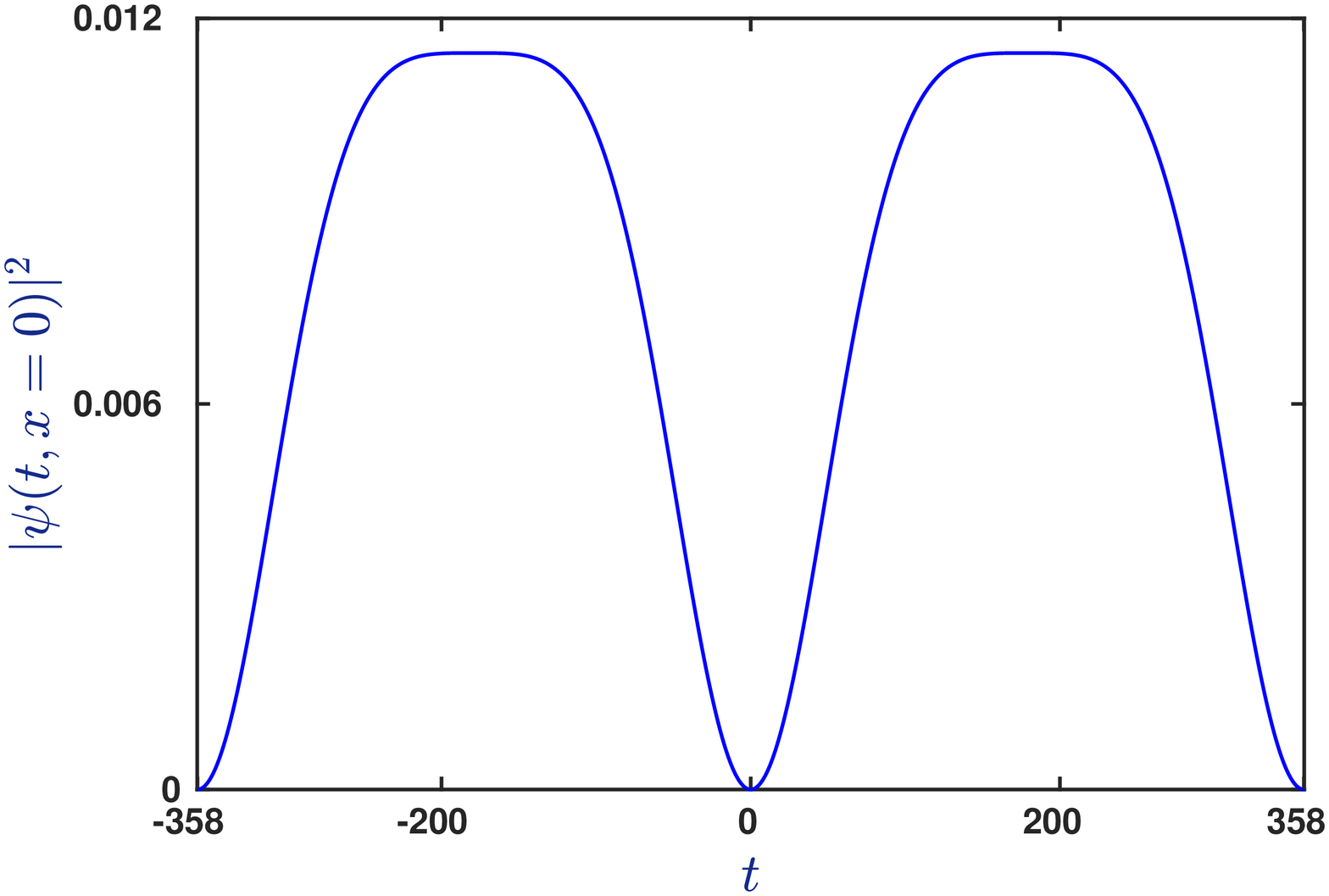}}
			~~
			\subfloat[\label{}]{\includegraphics[width=4.5cm,height=4.0cm]{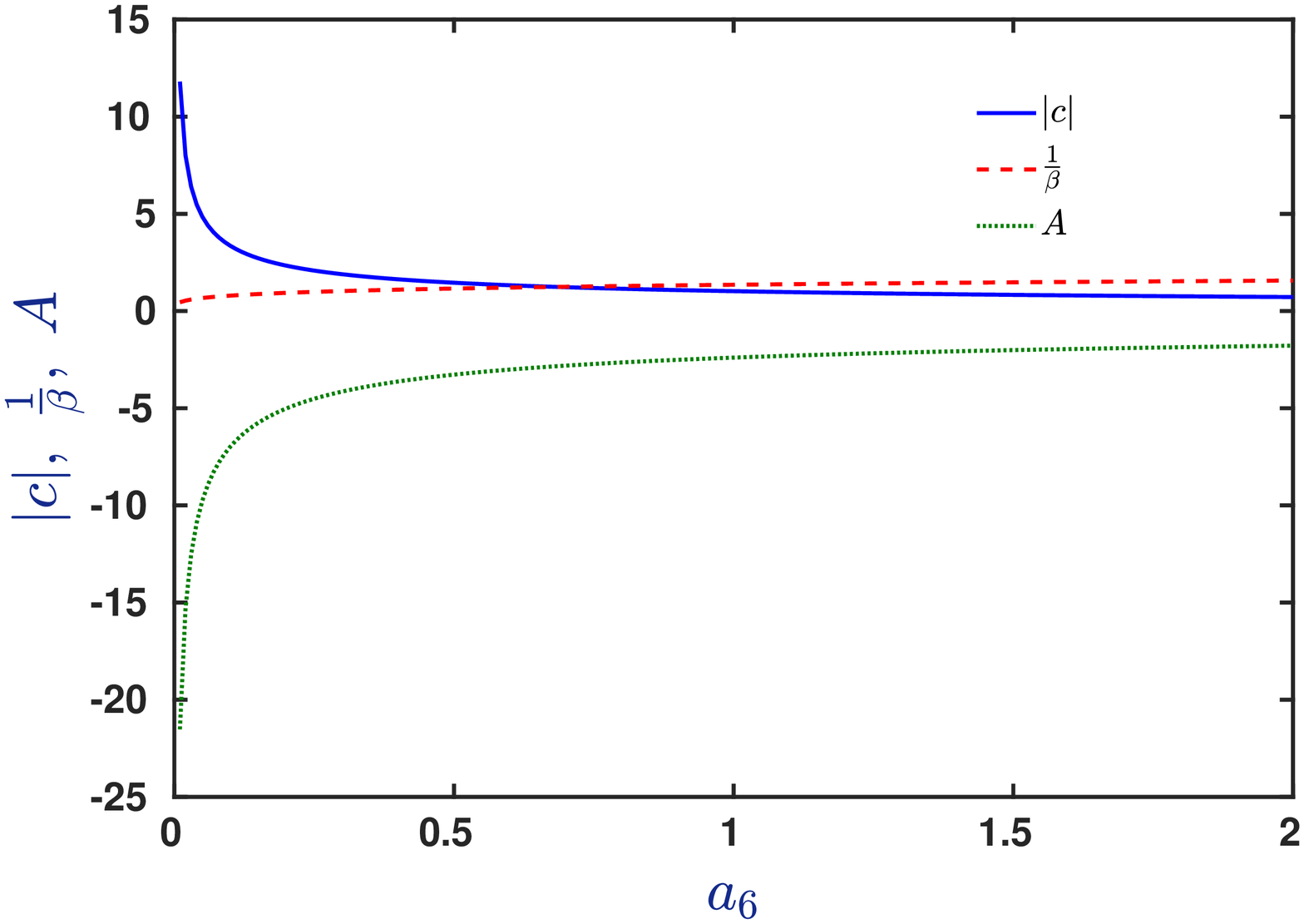}}
			~~
			\subfloat[\label{}]{\includegraphics[width=4.5cm,height=4.0cm]{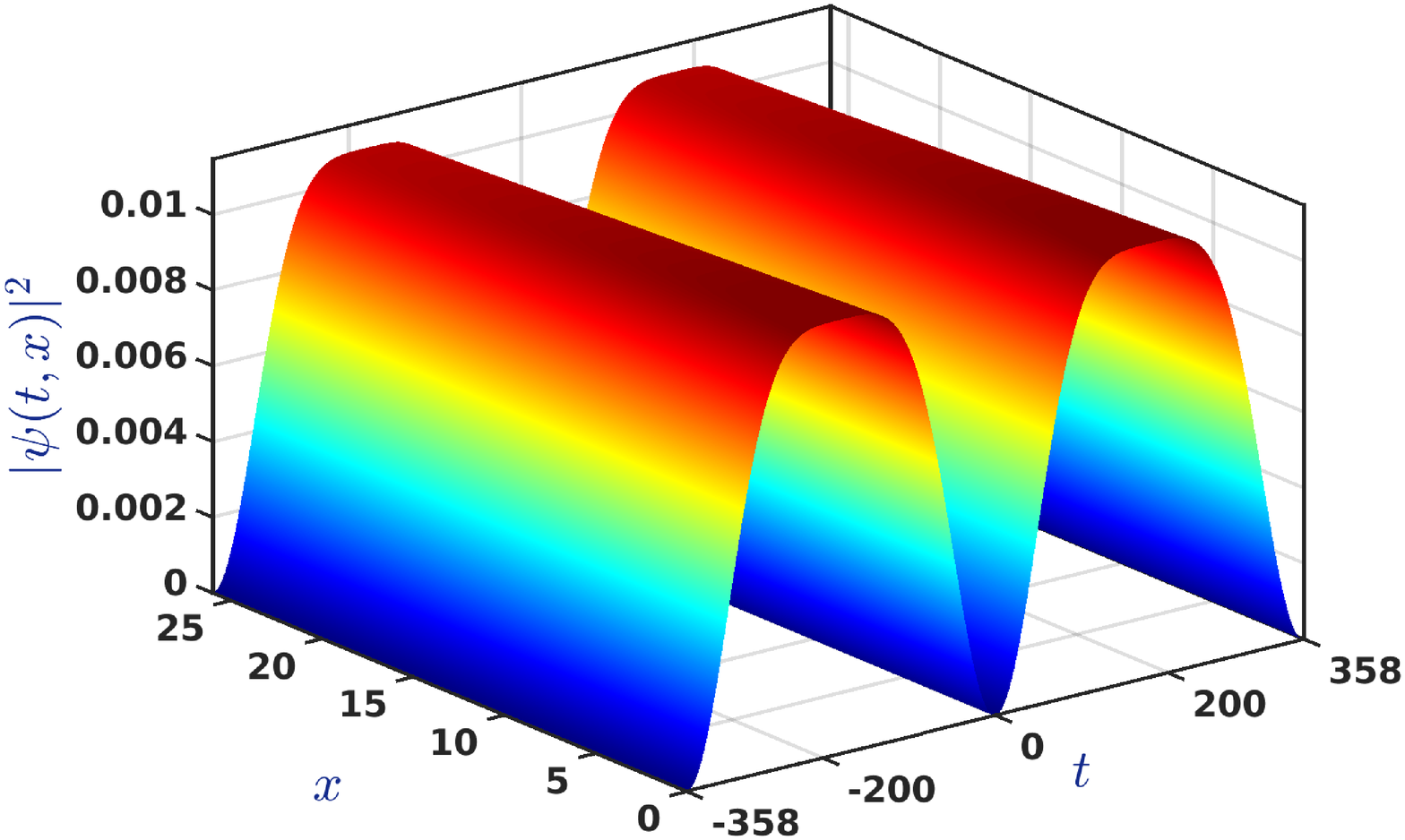}}\\
			\subfloat[\label{}]{\includegraphics[width=4.5cm,height=4.0cm]{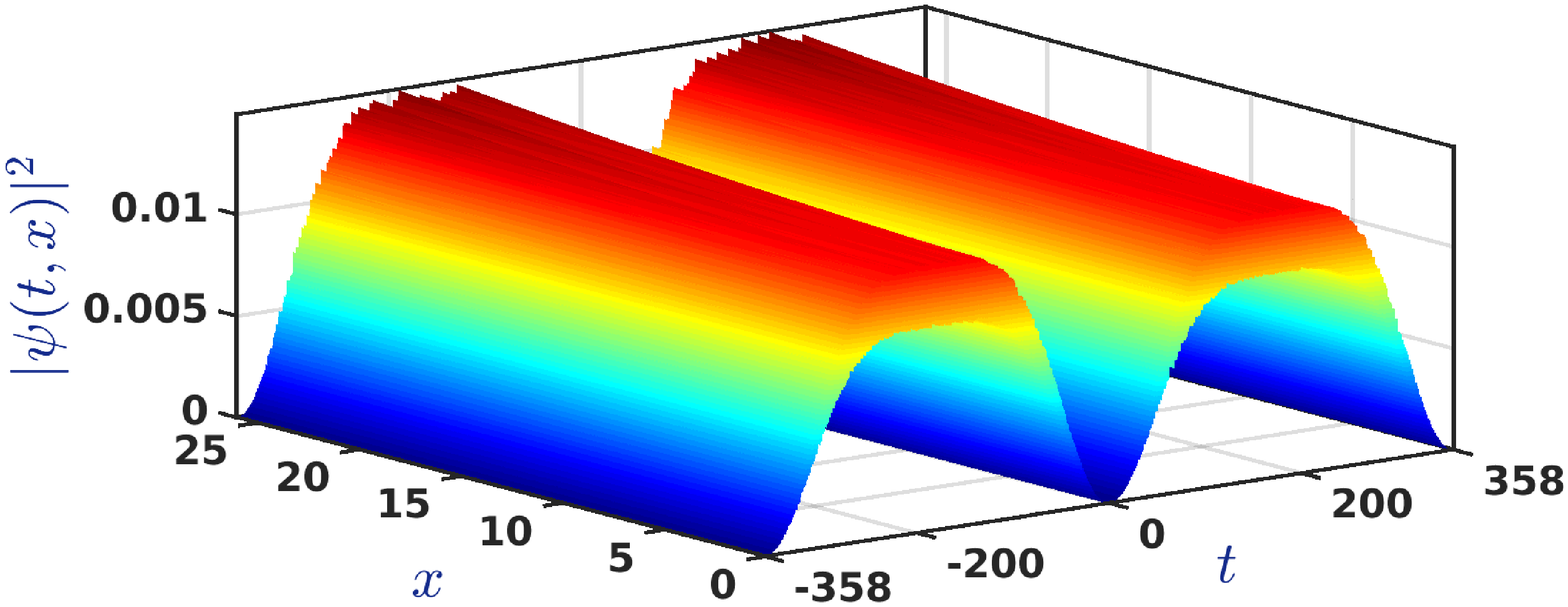}}
			~~
			\subfloat[\label{}]{\includegraphics[width=4.5cm,height=4.0cm]{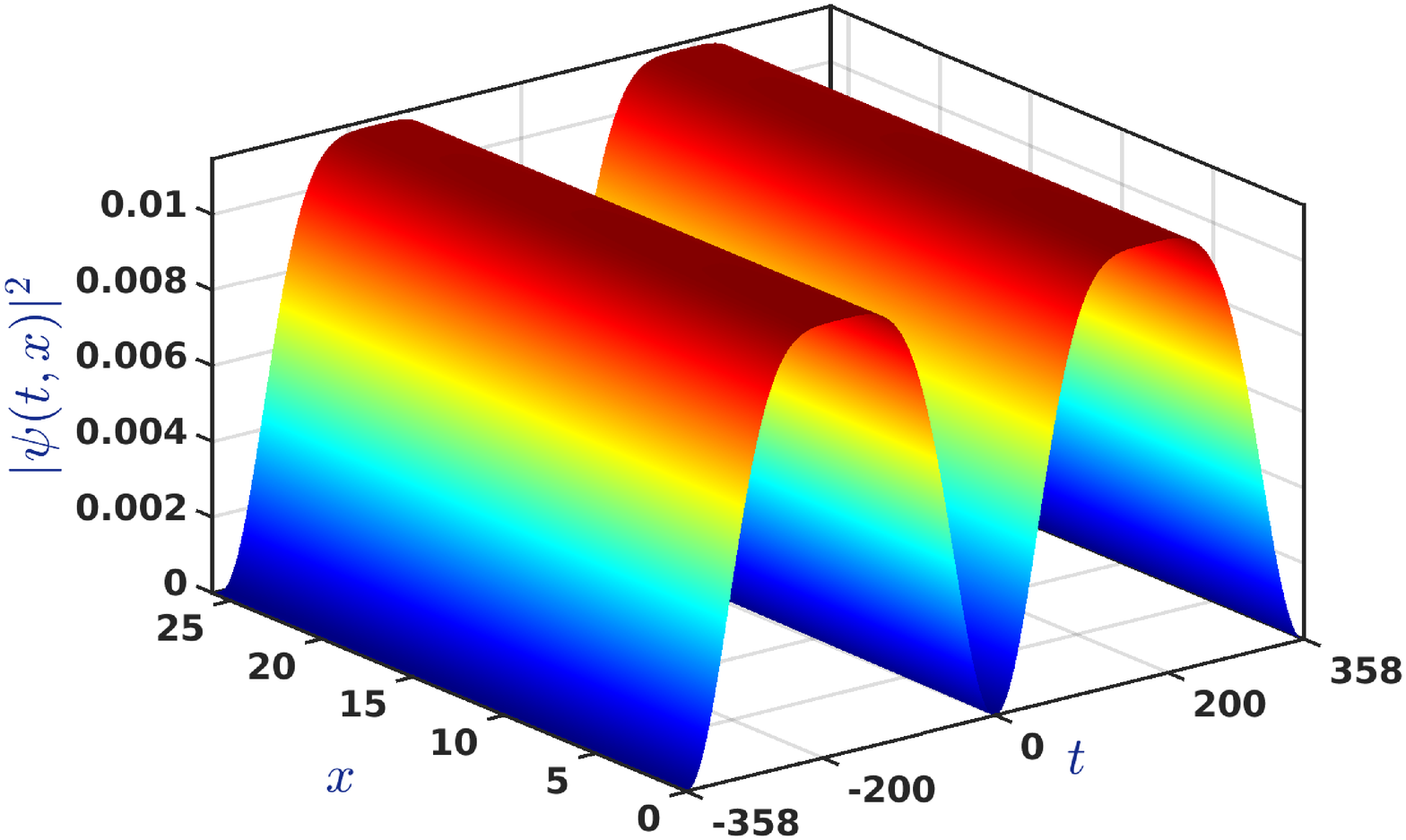}}
			~~
			\subfloat[\label{}]{\includegraphics[width=4.5cm,height=4.0cm]{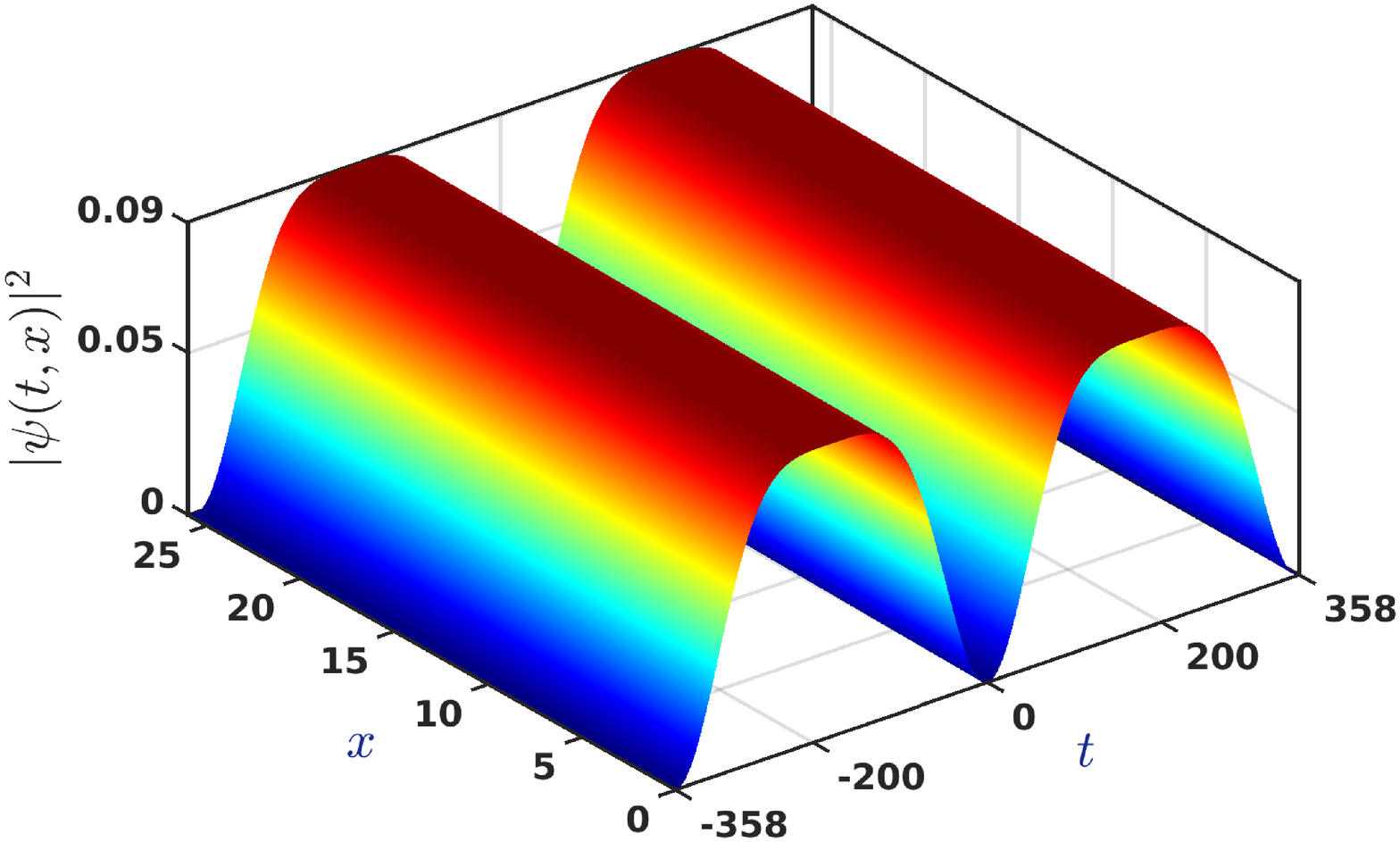}}
			\caption{(a) plot of $|\psi|^2$ versus $t$ at $x=0$ (c), (d) numerical simulation of the intensity profile without and with $10\%$ white noise respectively of the solution (\ref{1.20}) (dipole) for  $a_3=-0.165022,~a_4=0.1671553,~a_5=0.250733,~a_6=4.31016,~a_7=-0.006895,~\beta=0.0103625,~\omega=5.982461,~\phi_0=0.5,~c=0.476998,~k=1.301437,~A=0.2141115,~m=0.5$ respectively, (b) plots of $|c|$ , $\frac{1}{\beta}$, $A$  versus $a_6$ for the solution (\ref{1.20}) for $a_3=0.4,~a_4=0.5,~a_7=-0.04,~m=0.4$, (e) plot of numerical simulation of the intensity profile by adding $0.003$ to the parameters $a_i, i = 1,2, \cdots 7$ and (f) plot of  numerical simulation of the intensity profile by adding the overall factor $0.001$ to the parametric conditions (\ref{6}) \& (\ref{1.21}) respectively.}
		\end{center}
	\end{figure}
	
	\begin{figure}[]
		\begin{center}
		\subfloat[\label{}]{\includegraphics[width=4.5cm,height=4.0cm]{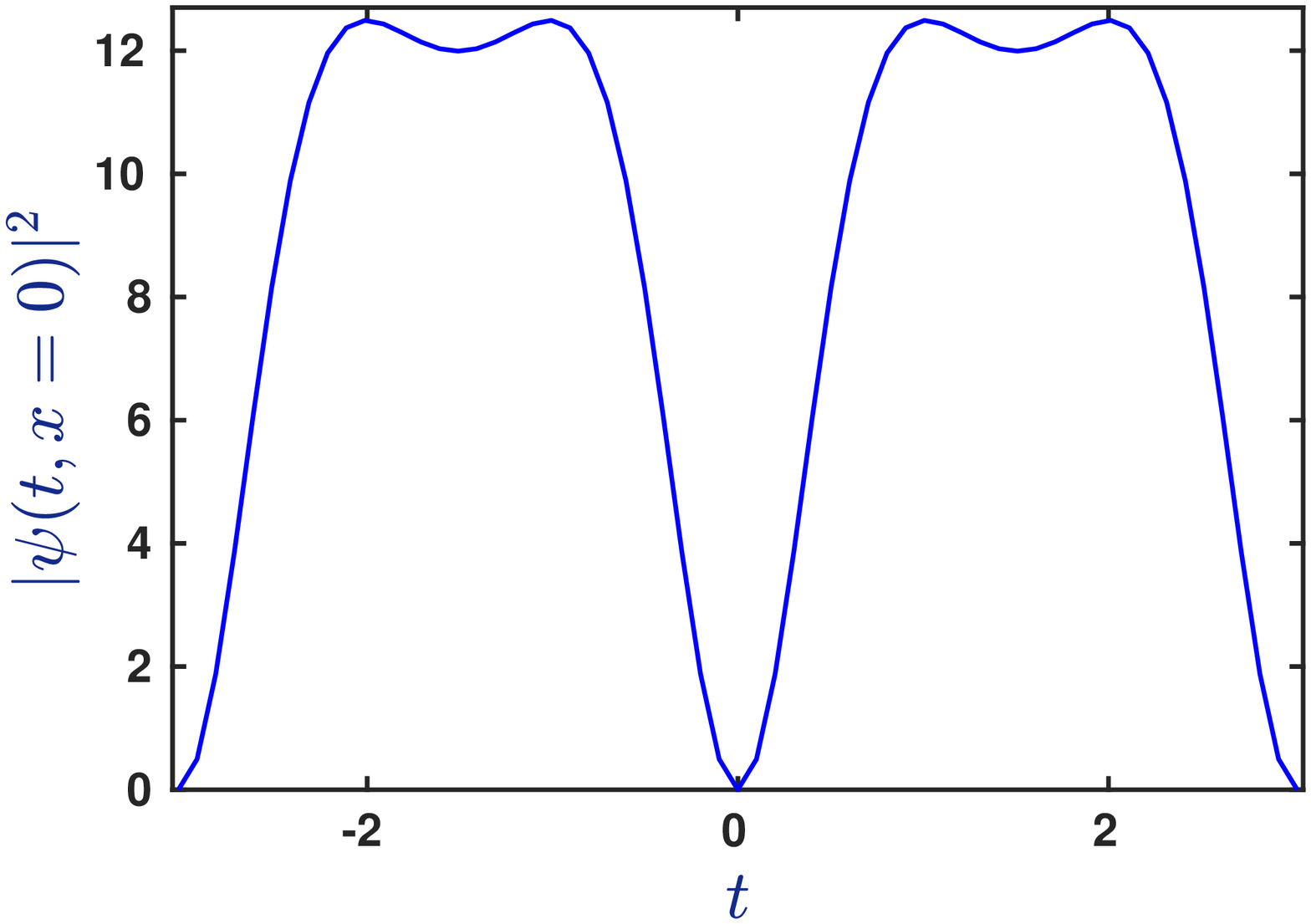}}
		~~
		\subfloat[\label{}]{\includegraphics[width=4.5cm,height=4.0cm]{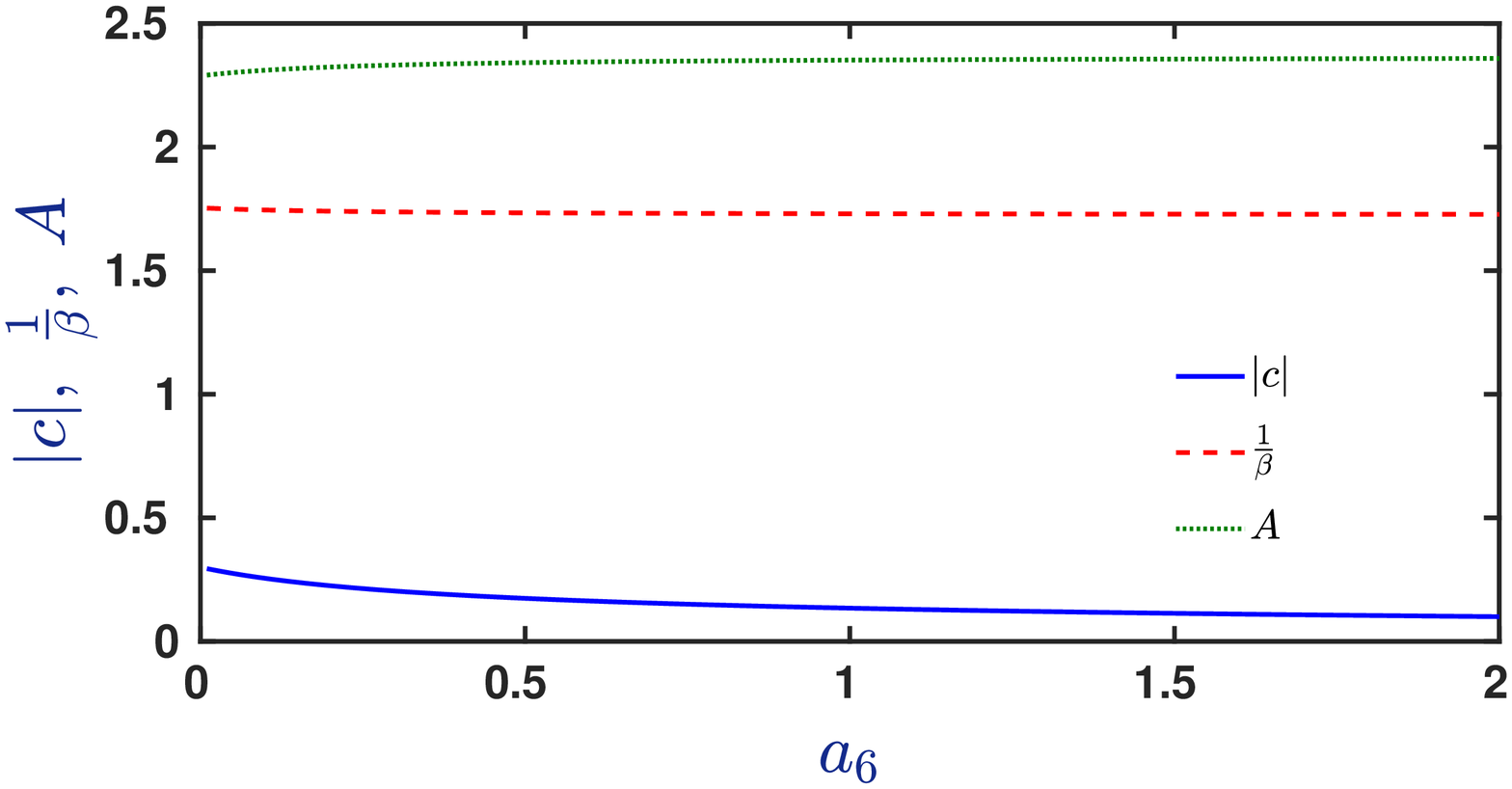}}
		~~
		\subfloat[\label{}]{\includegraphics[width=4.5cm,height=4.0cm]{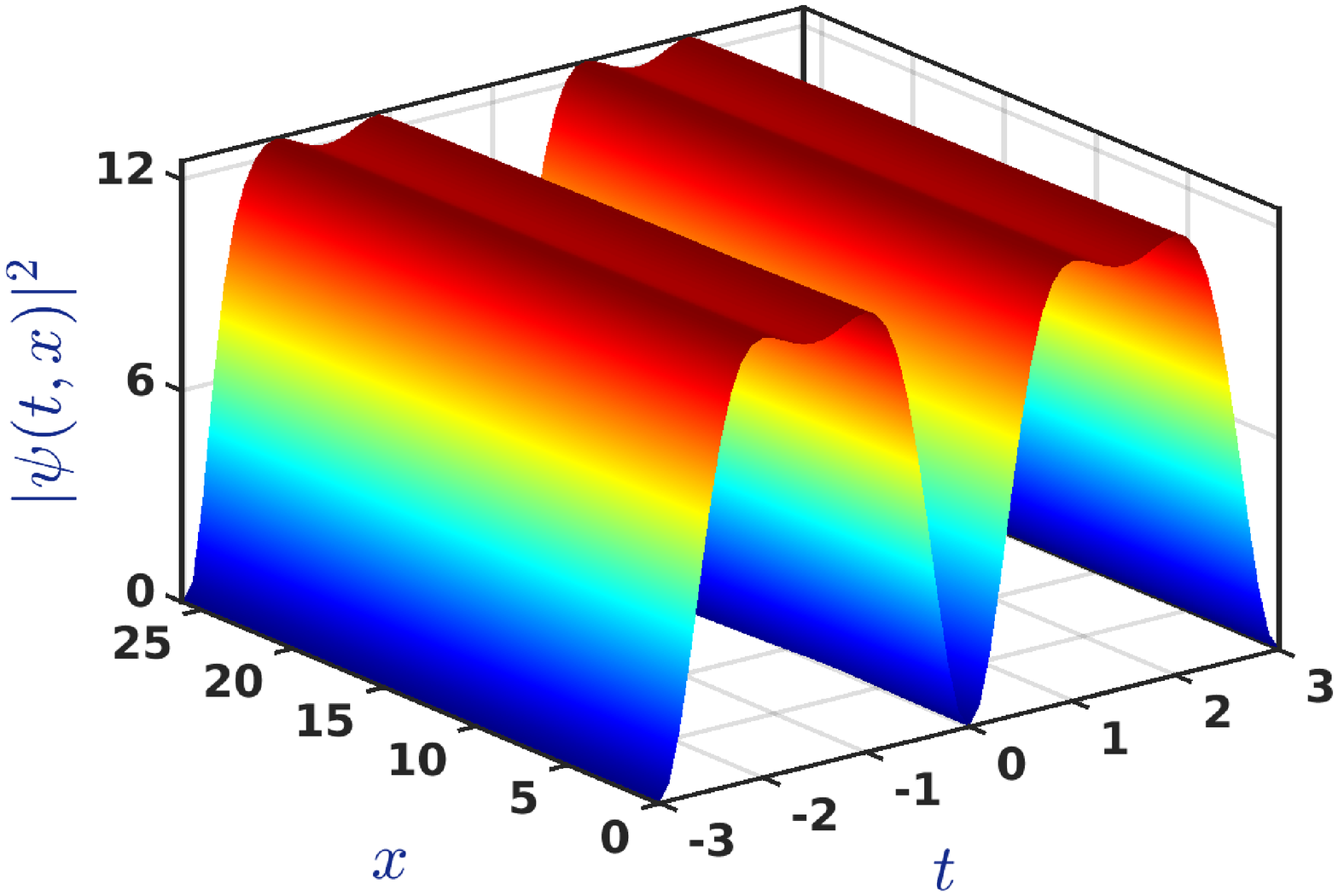}}\\
		\subfloat[\label{}]{\includegraphics[width=4.5cm,height=4.0cm]{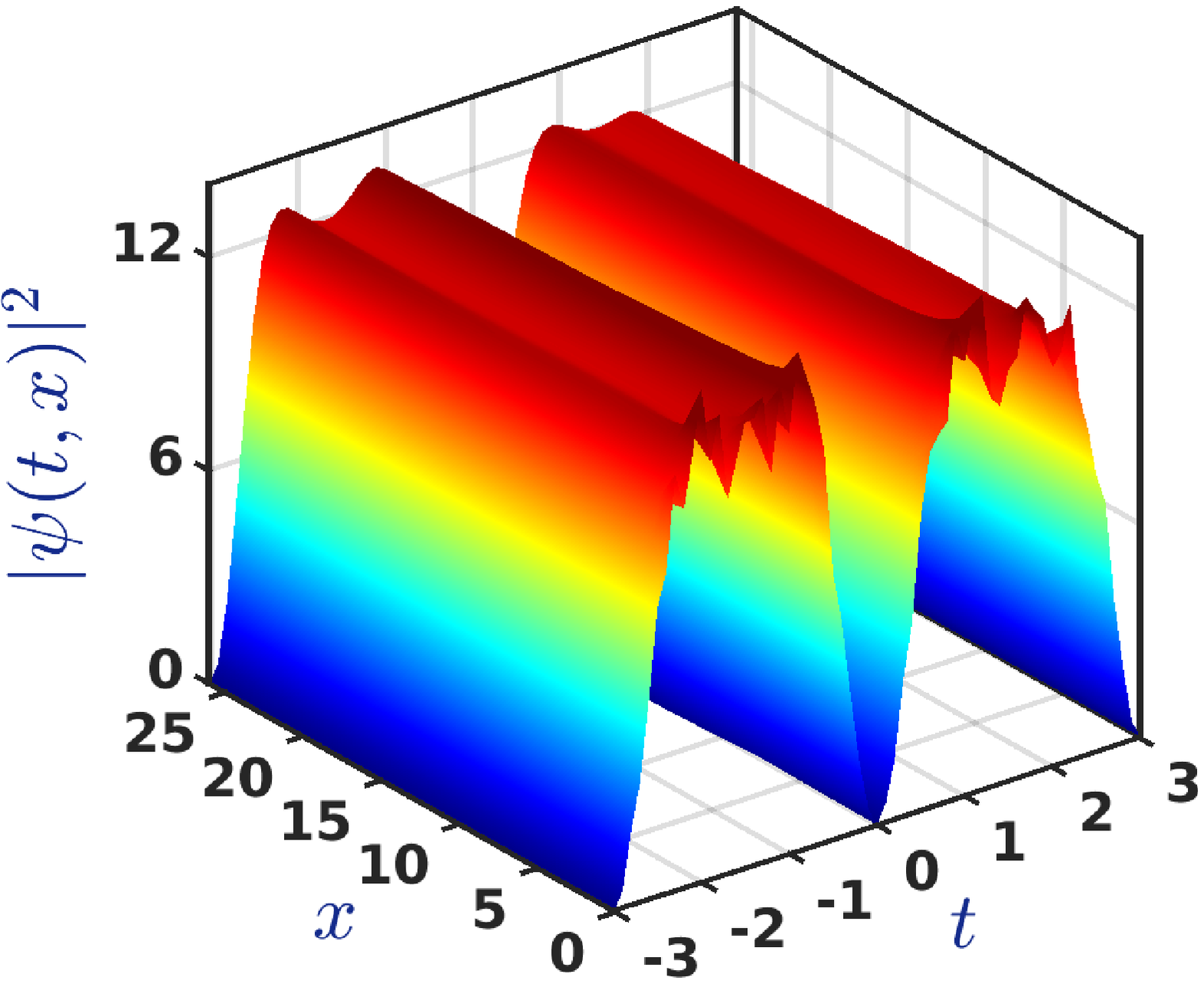}}
		~~
		\subfloat[\label{}]{\includegraphics[width=4.5cm,height=4.0cm]{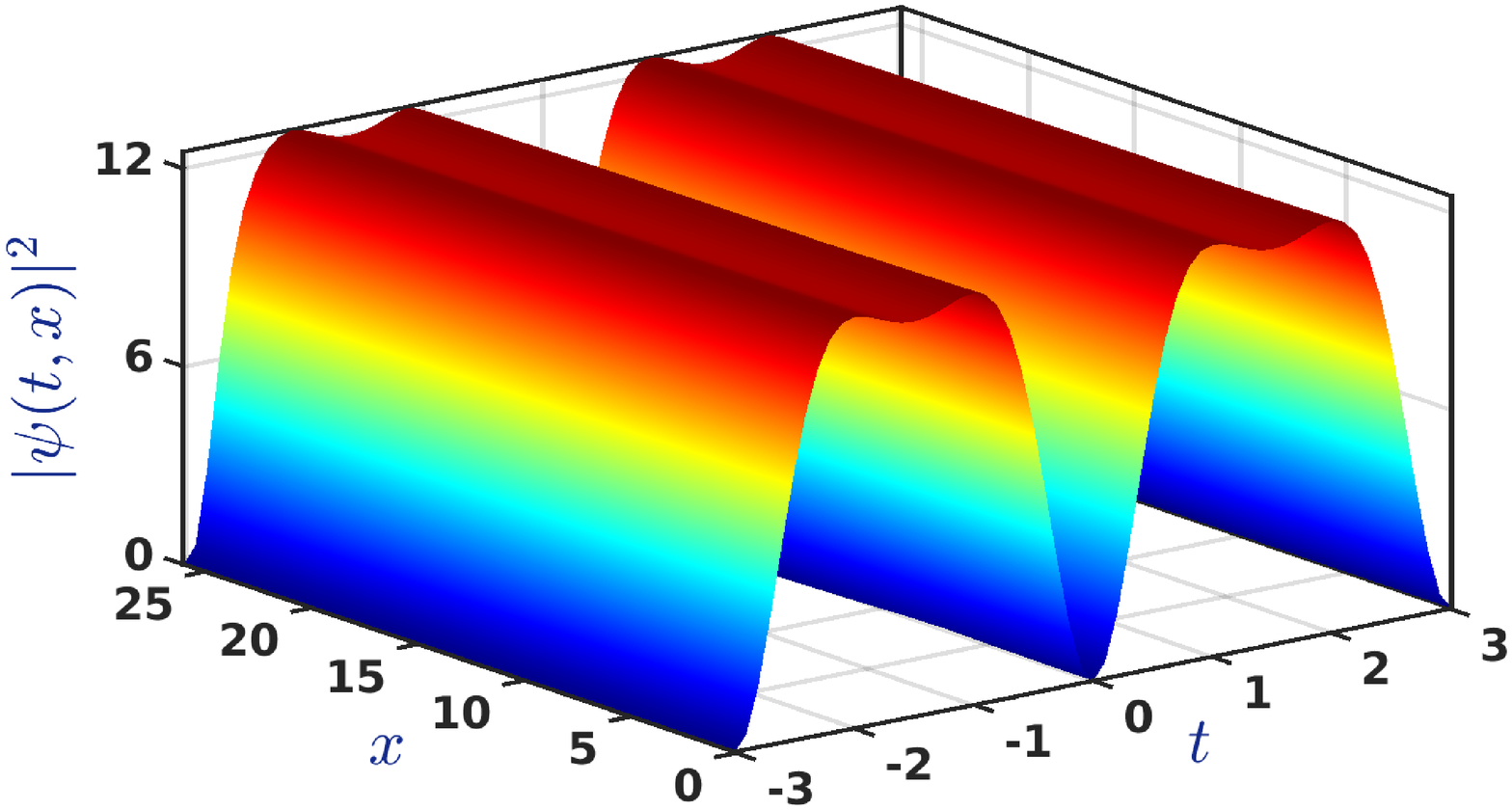}}
		~~
		\subfloat[\label{}]{\includegraphics[width=4.5cm,height=4.0cm]{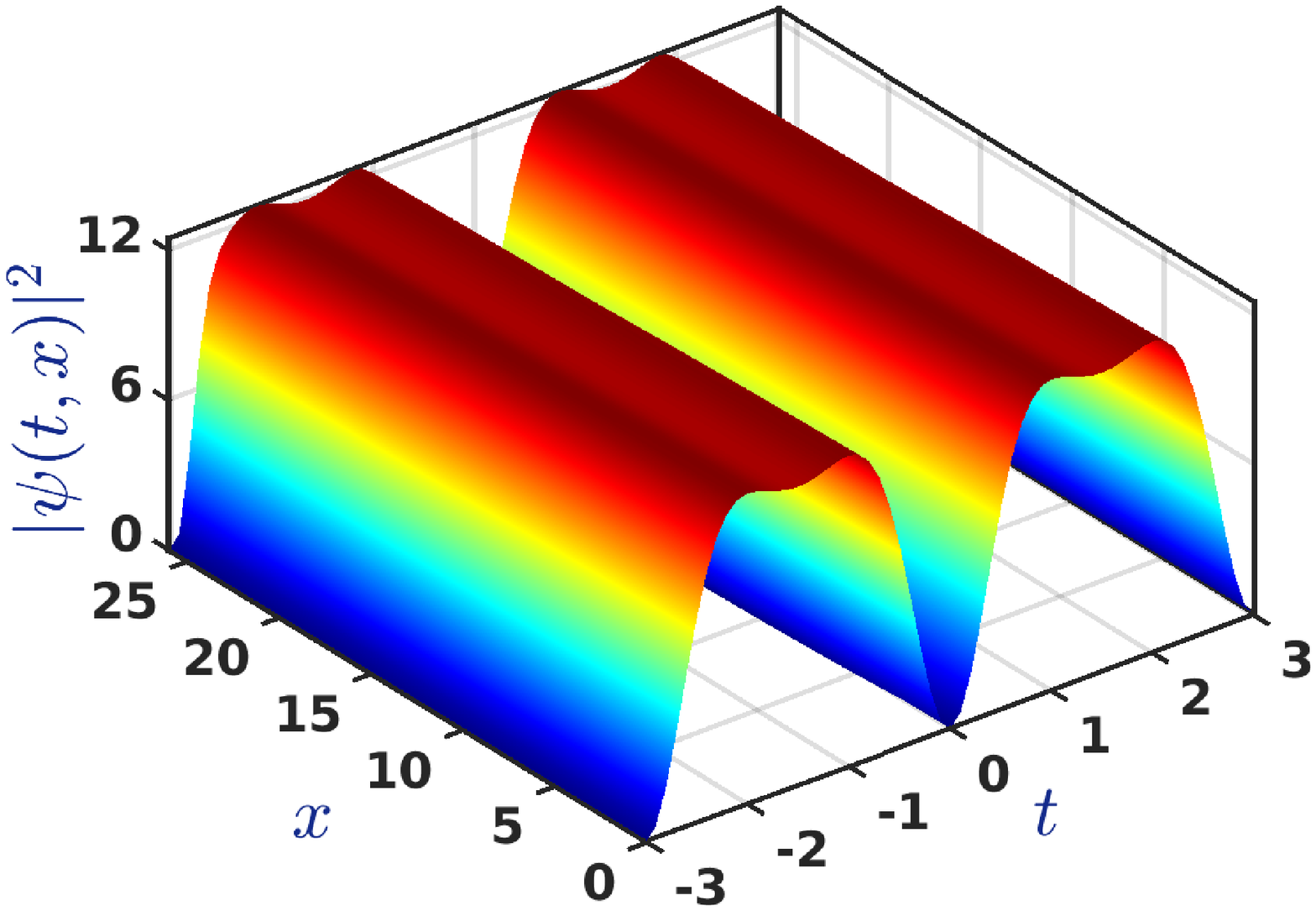}}
			\caption{(a) plot of $|\psi|^2$ versus $t$ at $x=0$ (c), (d) numerical simulation of the intensity profile without and with $10\%$ white noise respectively of the solution (\ref{1.20}) (quadrupole) for  $a_3=\frac{1}{60},~a_4=0.126,~a_5=-0.189,~a_6=0.95,~a_7=0.15,~\beta=1.291913,~\omega=0.027778,~\phi_0=0.5,~c=-0.004247,~k=2.919107,~A=7.0688,~m=0.6$ respectively, (b) plots of $|c|$, $\frac{1}{\beta}$ , $A$  versus $a_6$ of the solution (\ref{1.20}) for $a_3=0.4,~a_4=0.126,~a_7=0.4,~m=0.75$, (e) plot of numerical simulation of the intensity profile by adding $0.003$ to the parameters $a_i, i = 1,2, \cdots 7$ and (f) plot of numerical simulation of the intensity profiles by adding the overall factor $0.001$ to the parametric conditions (\ref{6}) \& (\ref{1.21}) respectively.}
		\end{center}
	\end{figure}
	
	{\bf Solution III. Periodic wave exhibiting dipole structure within a period}\\ \\
	Eqns.(\ref{1.5}) and (\ref{1.6}) admits yet another elliptic solitary wave 
	\begin{equation}\label{1.23}
		u(\xi) = A m sn(\xi, m) cn(\xi, m)\,,
	\end{equation}
($m, 0<m <1$ being the modulus of the Jacobi elliptic functions $sn, cn$, \cite{abra})	provided Eqs.(6a) - (6d) are satisfied and further the following relations are also satisfied 
	\begin{equation}\label{1.24}
		\begin{array}{lcl}
			\frac{1}{2}a_1 + c^2 a_6 = -2[3\omega^2-5(2-m) \beta^2] a_7\\
			k + k^2 a_6 + \frac{1}{2}\omega^2a_1  = [(11m^2+64m -64) \beta^4 - 3\omega^4] a_7
		\end{array}
	\end{equation}
	It is worth noting that when m=1, this periodic solution becomes the solution (\ref{1.12}). The intensity profile and numerical simulation of the intensity profile of the elliptic wave (\ref{1.23}) are shown in Figs.5(a) and (c) respectively. Stable evolution for the chosen values of the parameters are evident from the Fig. 5(c). From Figs.5(b), it is clear that the speed $|c|$ and the amplitude $A$ of this elliptic wave  decrease as $a_6$ increases  while the pulse-width $\frac{1}{\beta}$ increases as $a_6$ increases.\\
	
	\begin{figure}[]
		\begin{center}
		\subfloat[\label{}]{\includegraphics[width=4.5cm,height=4.0cm]{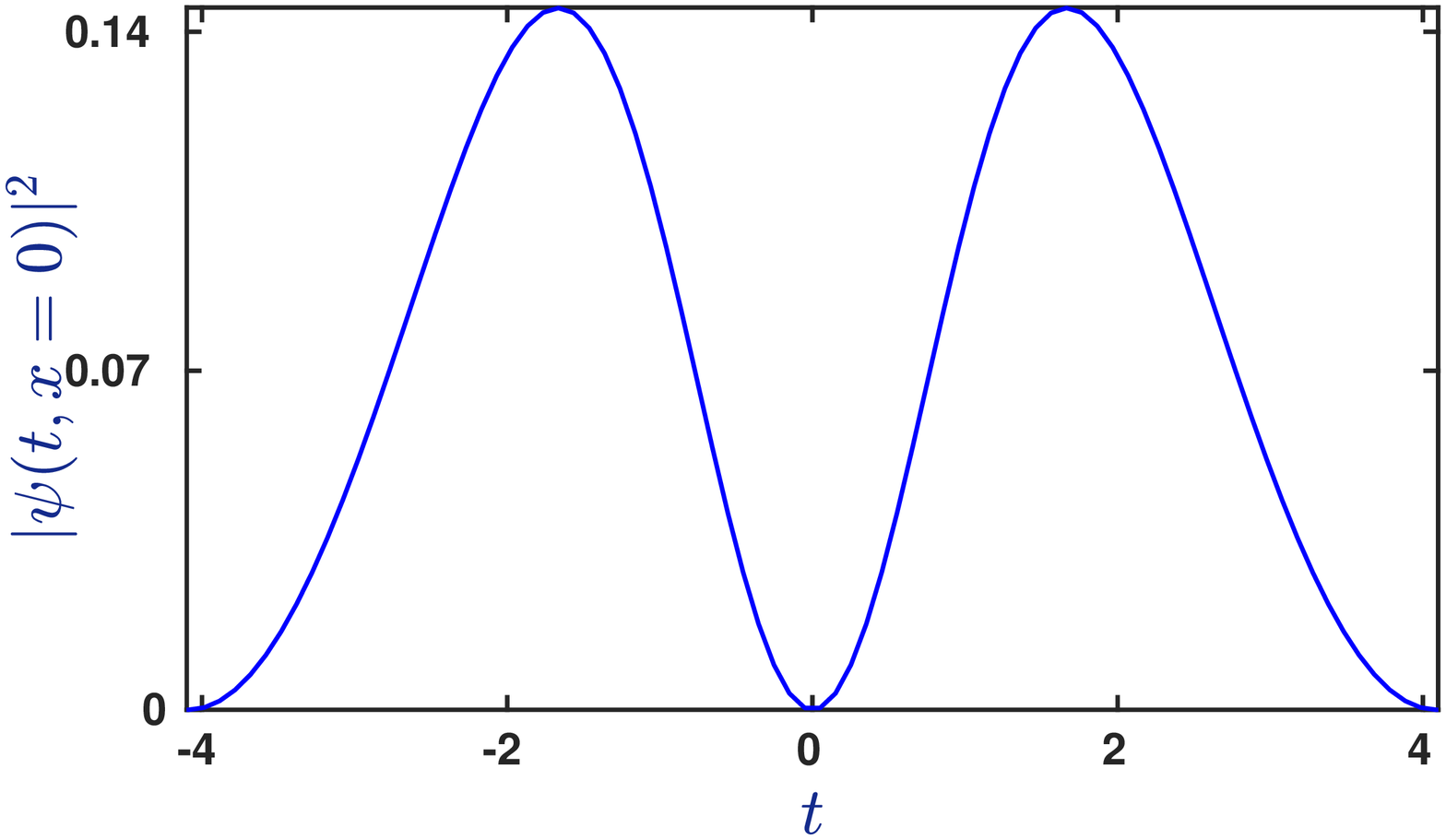}}
		~~
		\subfloat[\label{}]{\includegraphics[width=4.5cm,height=4.0cm]{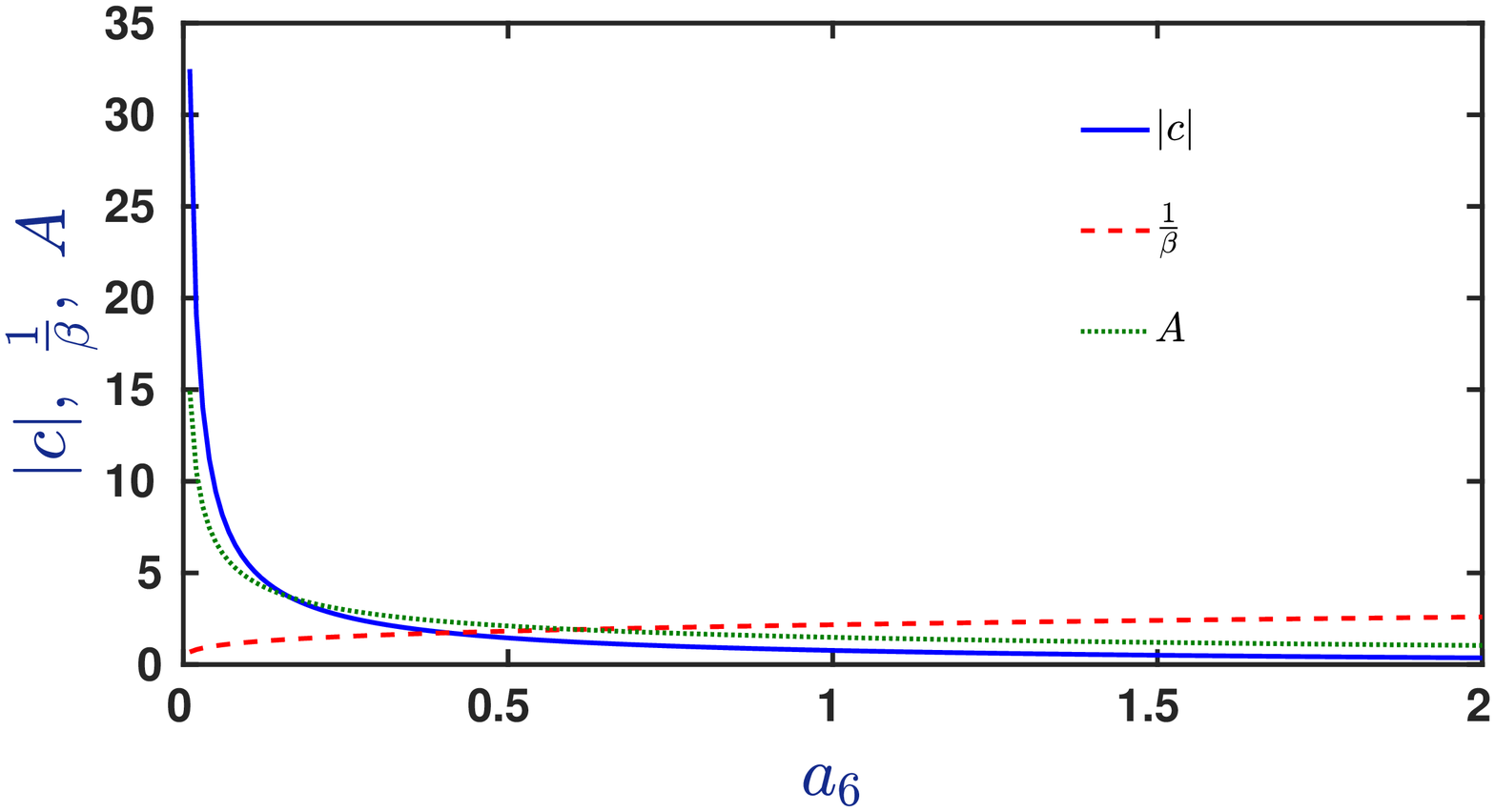}}
		~~
		\subfloat[\label{}]{\includegraphics[width=4.5cm,height=4.0cm]{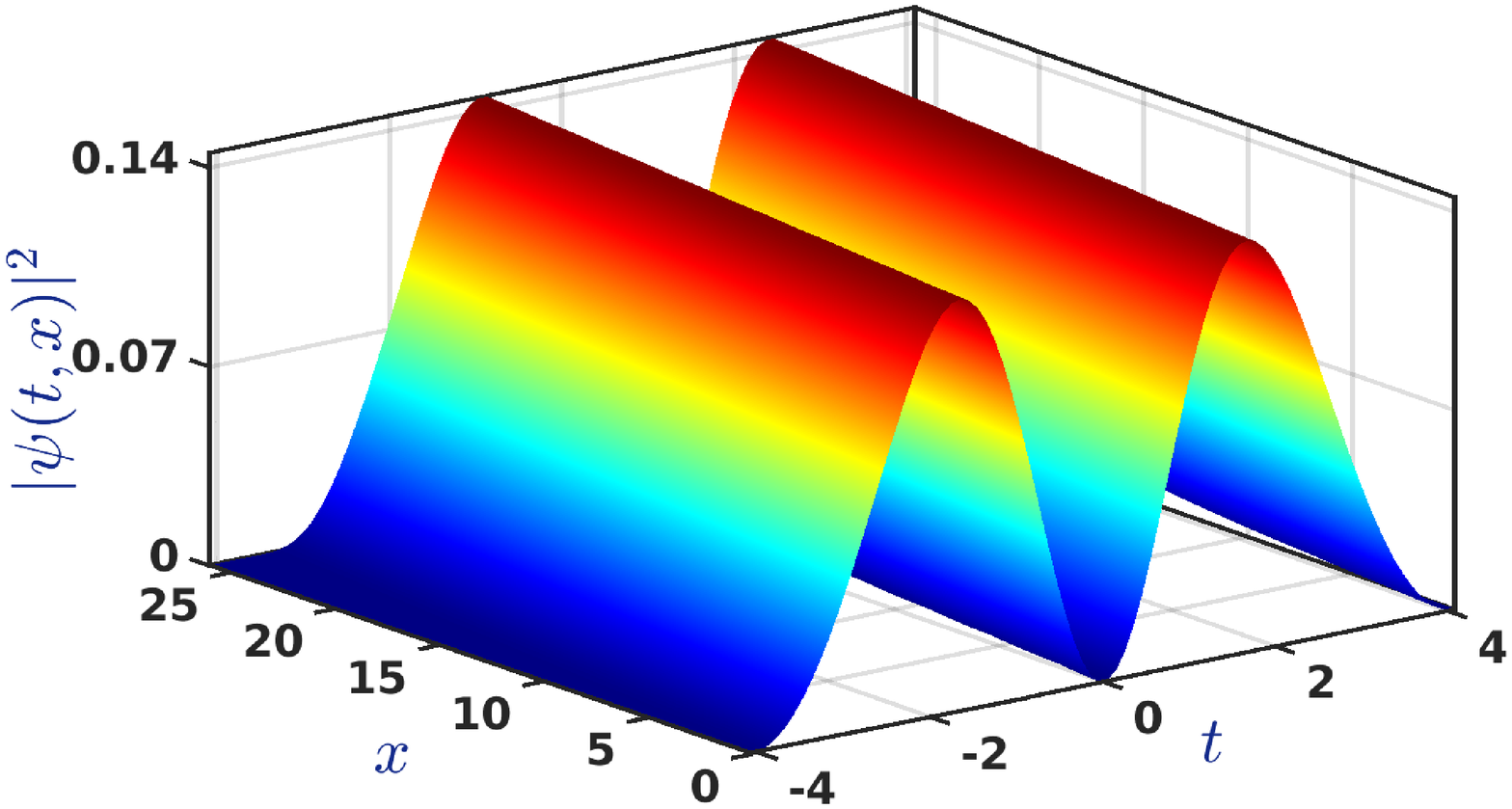}}\\
		\subfloat[\label{}]{\includegraphics[width=4.5cm,height=4.0cm]{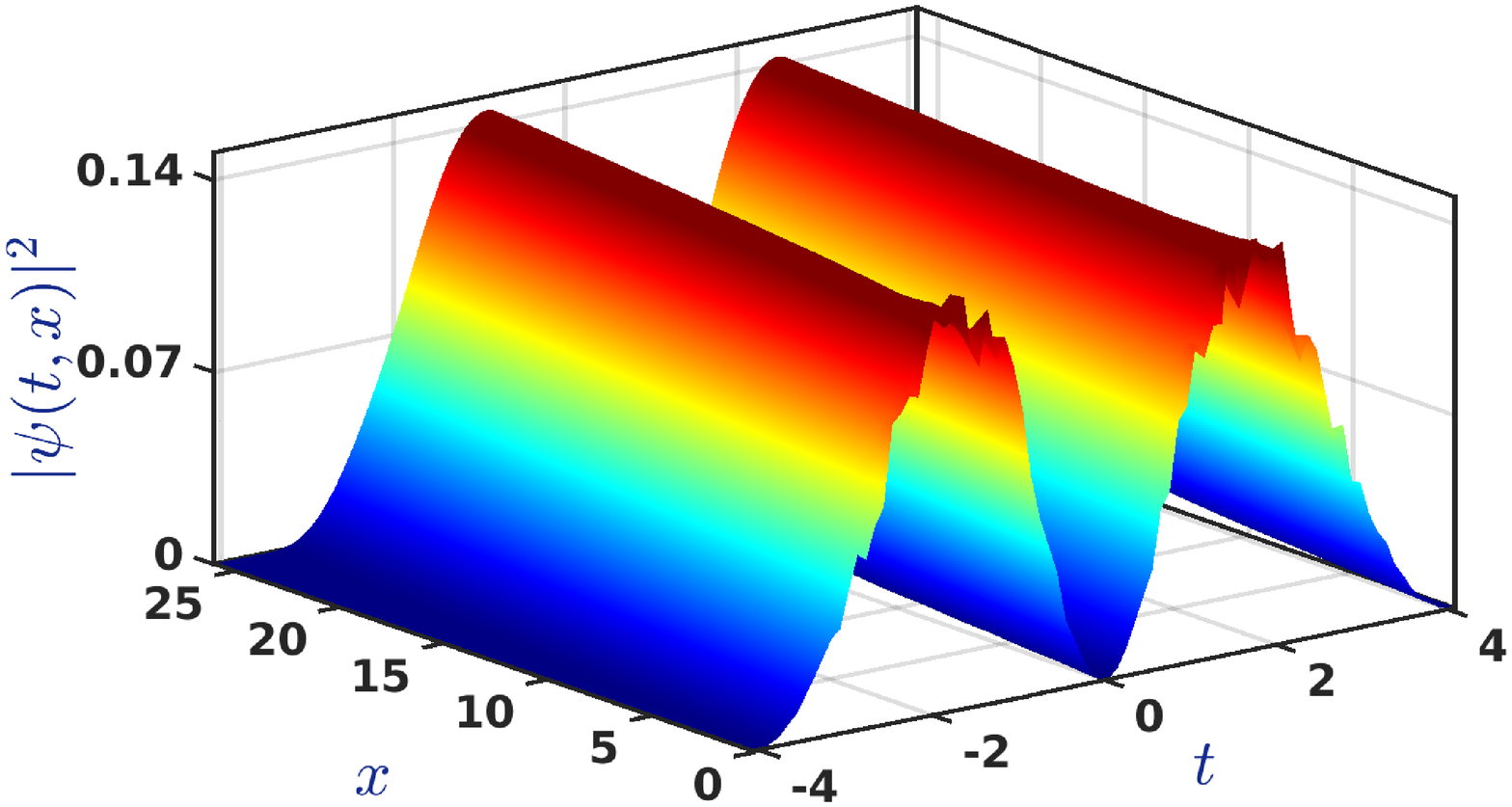}}
		~~
		\subfloat[\label{}]{\includegraphics[width=4.5cm,height=4.0cm]{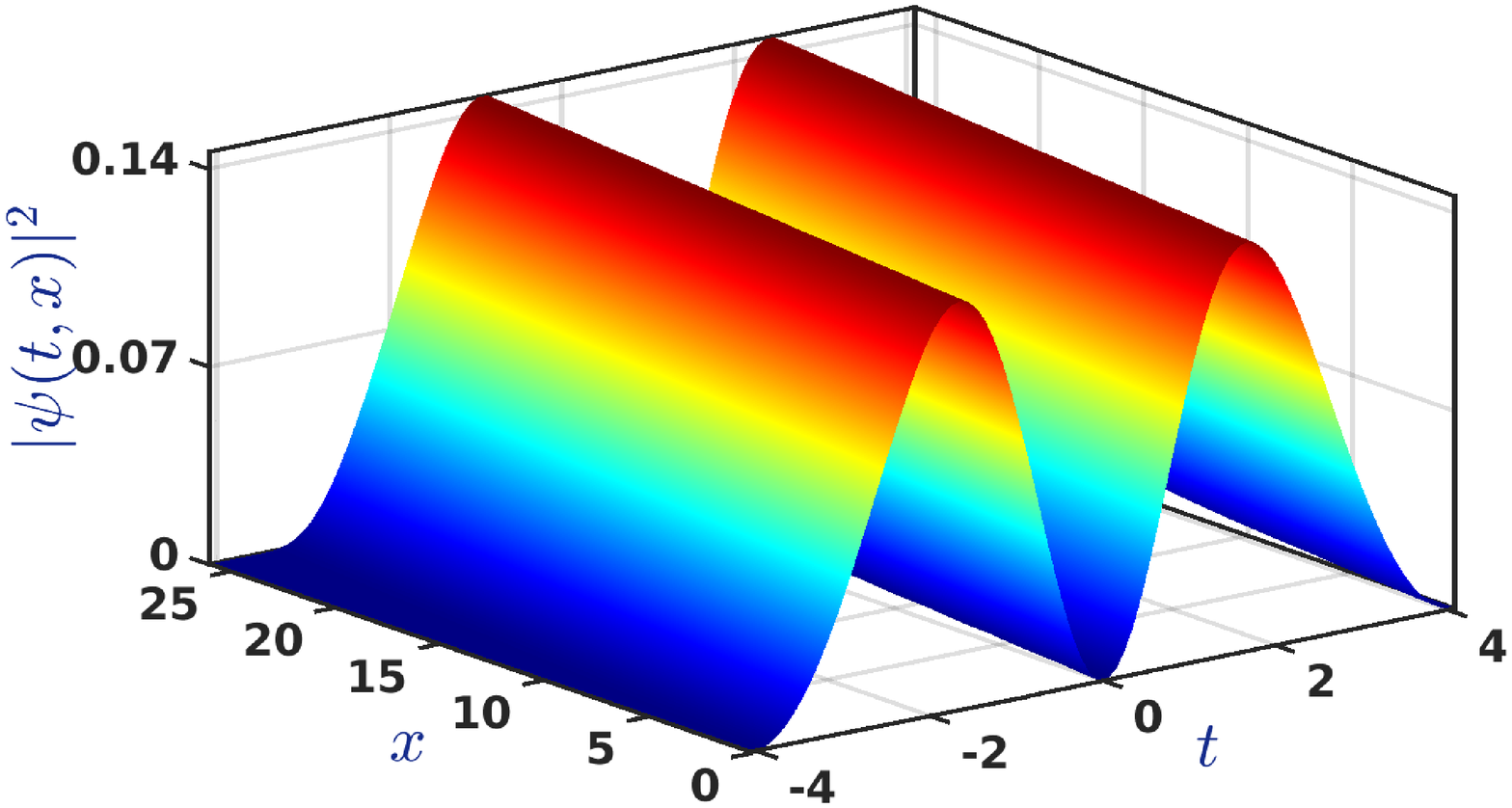}}
		~~
		\subfloat[\label{}]{\includegraphics[width=4.5cm,height=4.0cm]{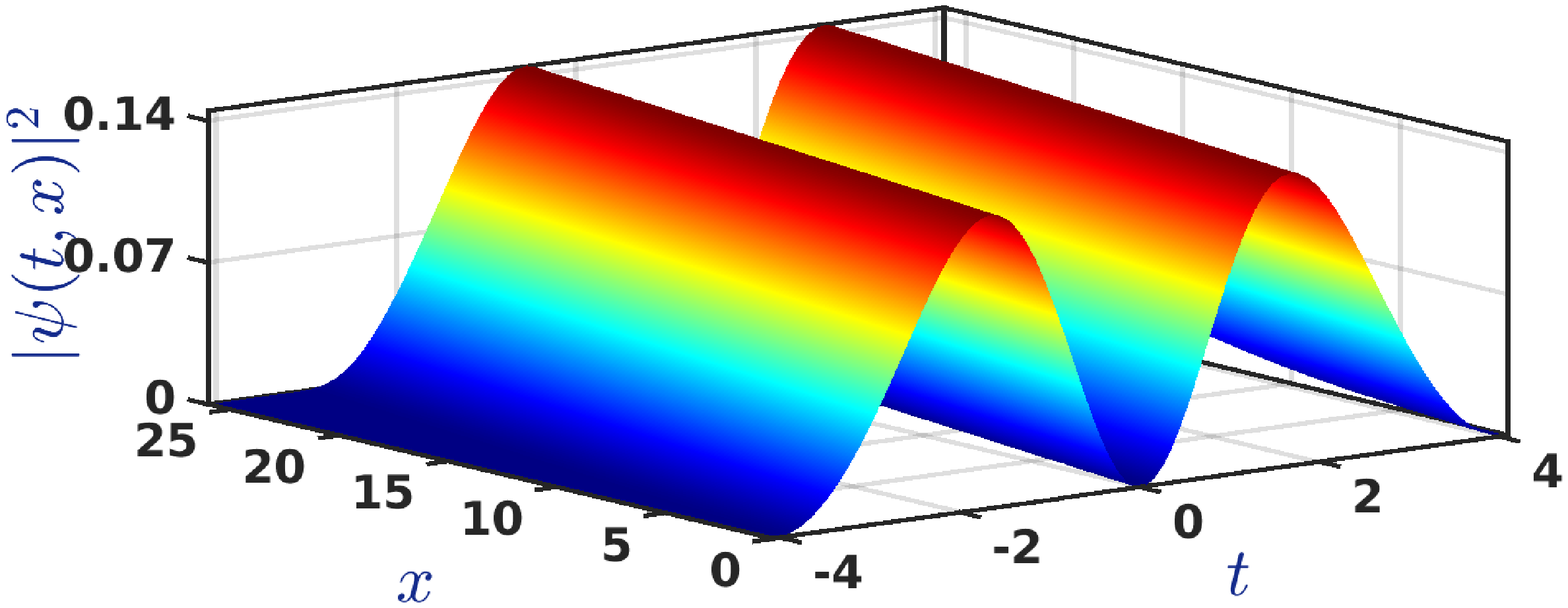}}
			\caption{(a) plot of $|\psi|^2$ versus $t$ at $x=0$ (c), (d) numerical simulation of the intensity profile without and with $10\%$ white noise respectively of the solution (\ref{1.23}) for  $a_3=\frac{1}{60},~a_4=0.126,~a_5=-0.189,~a_6=0.1,~a_7=0.15,~\beta=0.506763,~\omega=0.027778,~\phi_0=0.5,~c=0.0285981,~k=-9.861067,~A=1.0876,~m=0.7$ respectively, (b) plots of $|c|$, $\frac{1}{\beta}$, $A$ versus $a_6$ of the solution (\ref{1.23}) for $a_3=0.8,~a_4=-0.5,~a_7=0.4,~m=0.7$, (e) plot of numerical simulation of the intensity profile by adding $0.003$ to the parameters $a_i, i = 1,2, \cdots 7$ and (f) plot of numerical simulation of the intensity profile by adding the overall factor $0.001$ to the parametric conditions (\ref{6}) \& (\ref{1.24}) respectively.}
		\end{center}
	\end{figure}

	{\bf Solution IV. Periodic wave with arbitrary amplitude having dipole structure within a period}\\
	Remarkably, Eqns.(\ref{1.5}) and (\ref{1.6}) also admit the following trigonometric solitary wave solution with arbitrary amplitude $A$
	\begin{equation}\label{1.25}
		u(\xi) = A {\rm sin}(\xi)\,,
	\end{equation}
	provided following relations are also satisfied 
	\begin{equation}\label{1.26}
		\begin{array}{lcl}
			3a_4+2a_5=0\\
			4(a_3-4\omega a_7)\beta^2+(1+2ka_6)c+(\omega a_1+3\omega^2a_3-4\omega^3a_7)=0\\
			\omega a_4+a_2=0\\
			a_7(16\beta^4+\omega^4+24\beta^2\omega^2)-a_3\omega(12\beta^2+\omega^2)-2\beta^2(a_1+2a_6c^2)-(k+k^2a_6+\frac{1}{2}\omega^2a_1)=0\\
		\end{array}
	\end{equation}
	The intensity profile and numerical simulation of the intensity profile of the solitary wave (\ref{1.25}) exhibiting stable evolution for the chosen parameters are shown in Figs.6(a) and (c) respectively. Fig. 6(b) shows that the 
	pulse-width $\frac{1}{\beta}$ increases as $a_6$ increases but $|c|$ decreases with increasing $a_6$.
	\begin{figure}[]
		\begin{center}
			\centering
			\subfloat[\label{}]{\includegraphics[width=4.5cm,height=4.0cm]{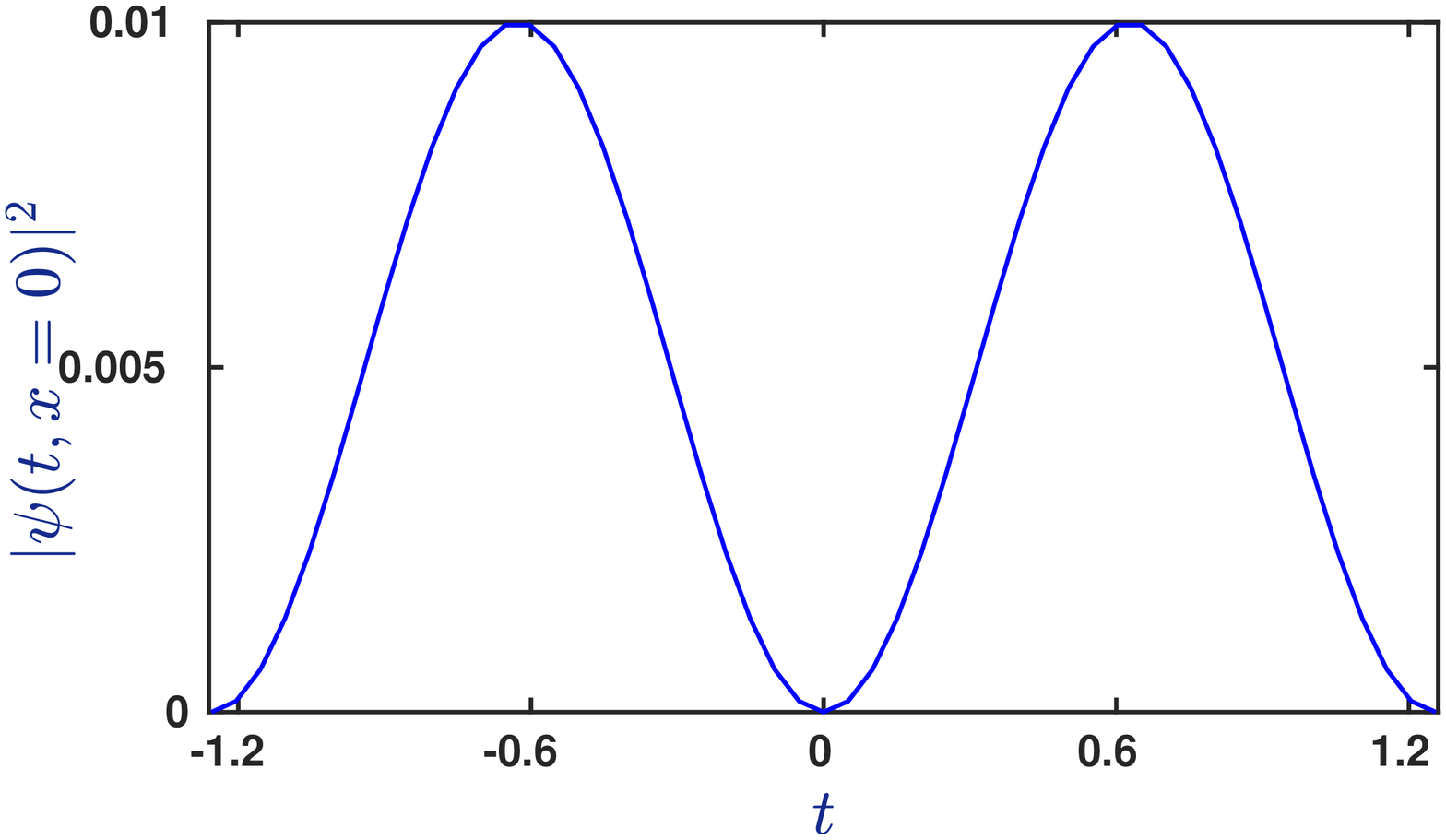}}
			~~
			\subfloat[\label{}]{\includegraphics[width=4.5cm,height=4.0cm]{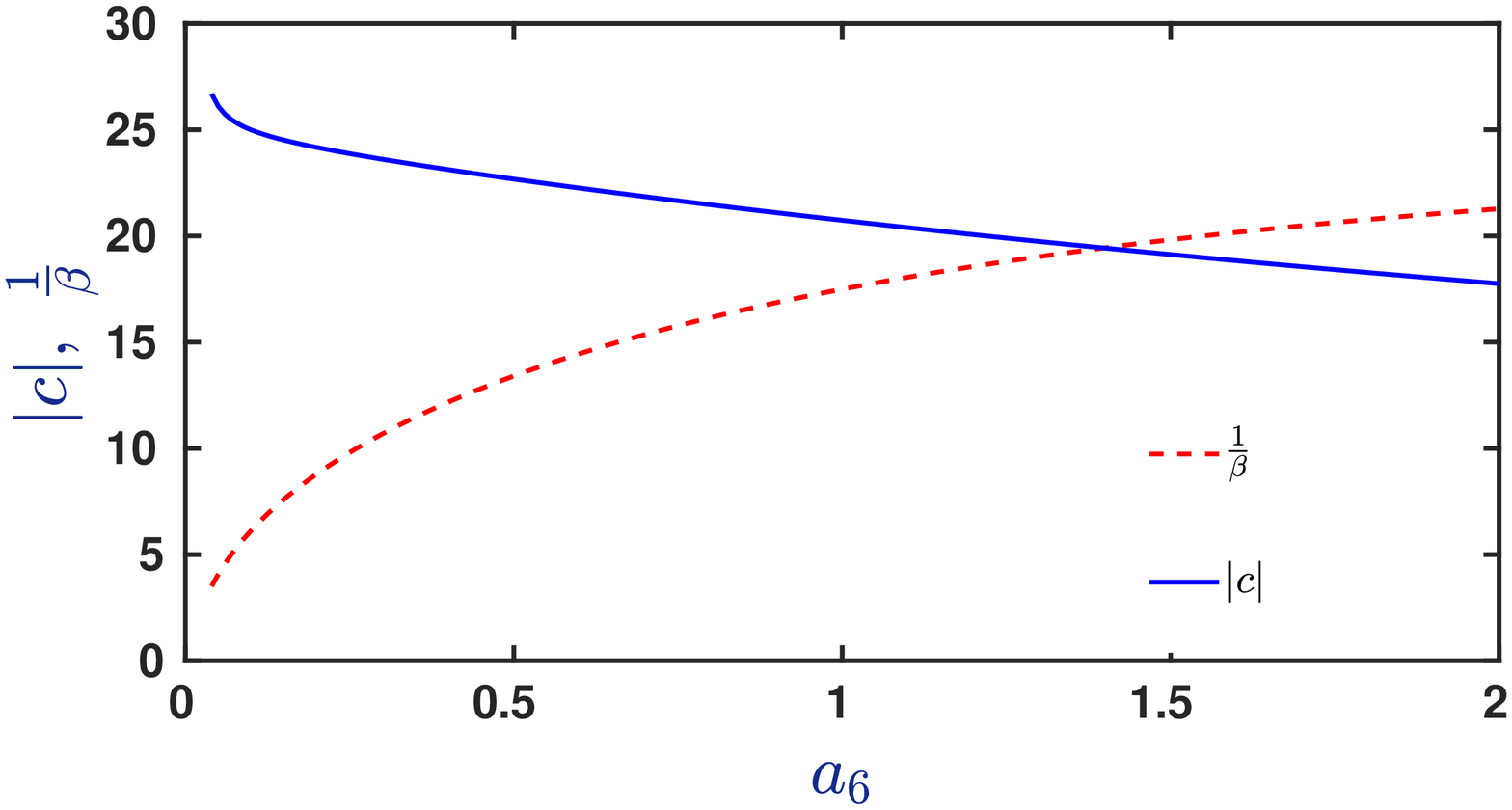}}
			\subfloat[\label{}]{\includegraphics[width=4.5cm,height=4.0cm]{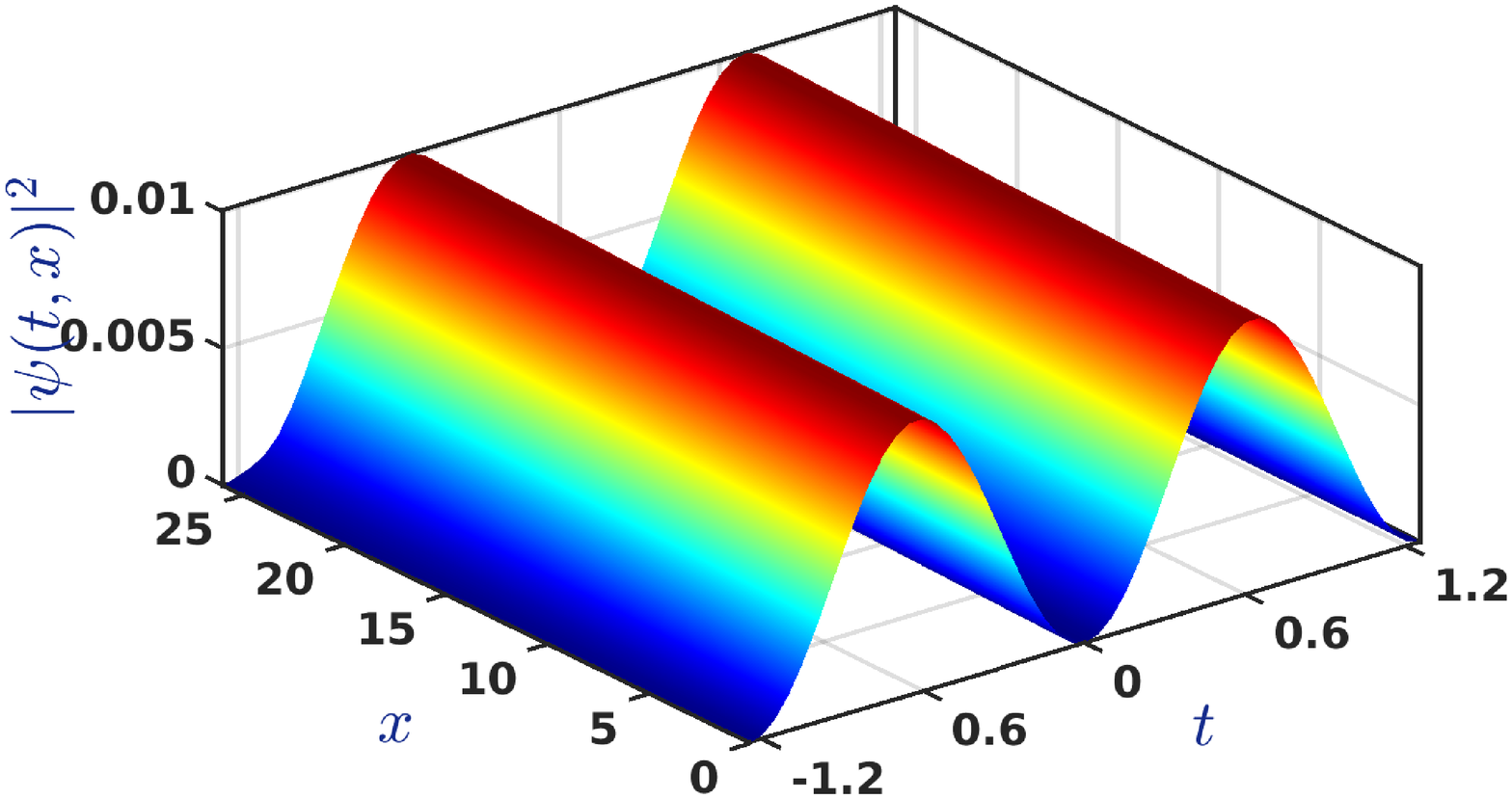}}\\
			\subfloat[\label{}]{\includegraphics[width=4.5cm,height=4.0cm]{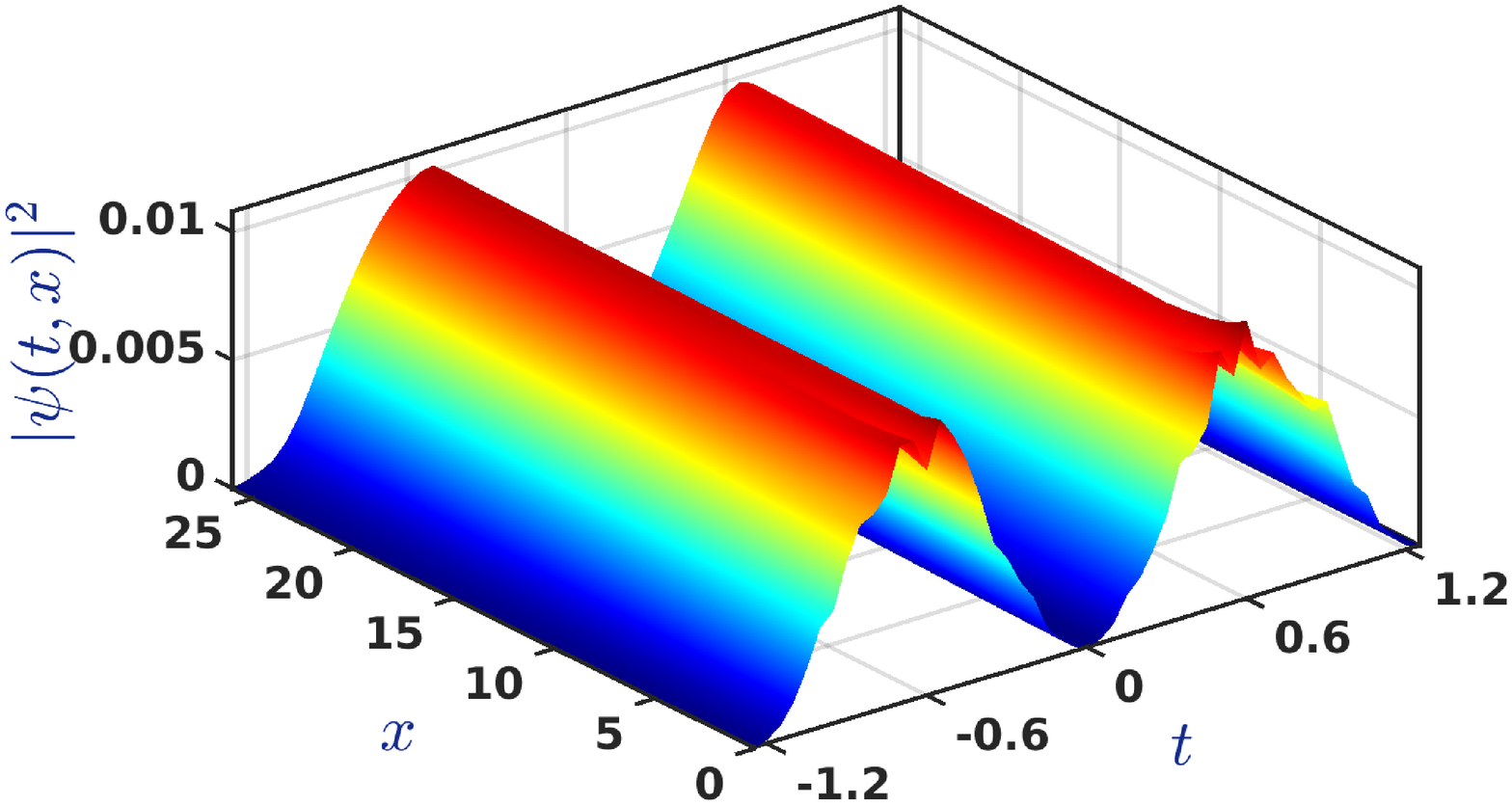}}
			~~
			\subfloat[\label{}]{\includegraphics[width=4.5cm,height=4.0cm]{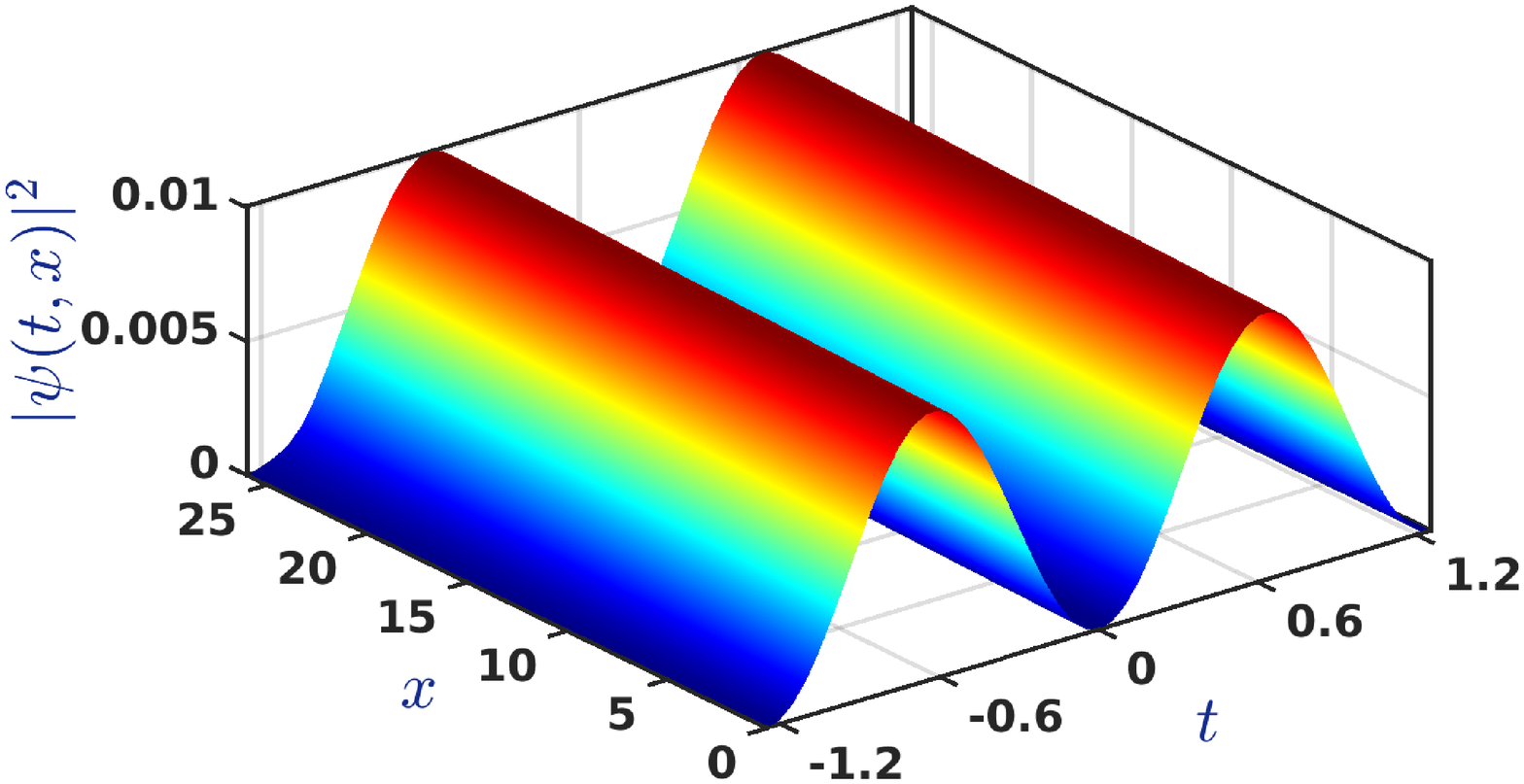}}
			~~
			\subfloat[\label{}]{\includegraphics[width=4.5cm,height=4.0cm]{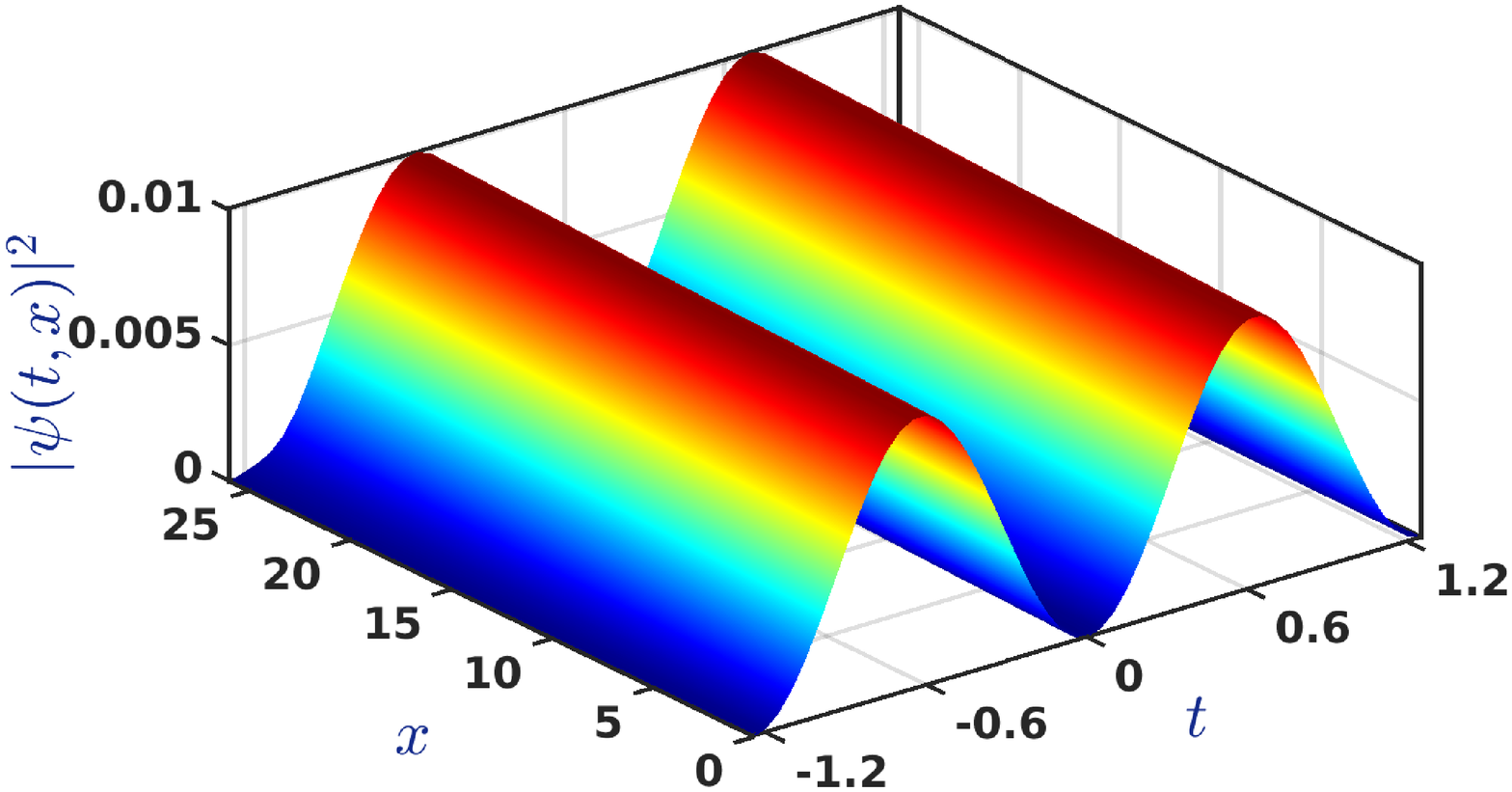}}
			\caption{(a) plot of $|\psi|^2$ versus $t$ at $x=0$ (c), (d) numerical simulation of the intensity profile without and with $10\%$ white noise respectively of the solution (\ref{1.25}) for  $a_3=0.1,~a_4=-0.4,~a_5=0.6,~a_6=0.5,~a_7=0.04,~\beta=1.25169,~\omega=2.5,~\phi_0=0.5,~c=0.005,~k=0.0125112,~A=0.1$ respectively, (b) plots of $|c|$,  $\frac{1}{\beta}$ versus $a_6$ respectively of the solution (\ref{1.25}) for $a_3=0.1,~a_4=0.5,~a_6=0.04,~a_7=0.8,~k=0.1$, (e) plot of numerical simulation of the intensity profile by adding $0.003$ to the parameters $a_i, i = 1,2, \cdots 7$ and (f) plot of numerical simulation of the intensity profile by adding the overall factor $0.001$ to the parametric conditions (\ref{6}) \& (\ref{1.26}) respectively.}
		\end{center}
	\end{figure}

	\section{Stability analysis}
	 The physical importance of the solitary wave solutions rests on their stability. In this regard, it is worth noting that Figures 2(c), 3(c), 4(c), 5(c), and 6(c) all refer to an ideal setting, which is not always the case in real-world situations. There are a variety of factors that might induce instability in propagating structures \cite{triki2019}. As a result, it becomes vital to investigate the stability of solutions in the presence of external noise or perturbations. For this, $10\%$ white noise perturbation is added to the initial input, 
	 as $\psi(x=0,t)=u(\beta t)e^{i\phi(x=0,t)}(1+\epsilon)$, where $u(\beta t)$ are exact solutions which are given by (8), (12), (14), (16)  and the white noise $\epsilon$ is realized by utilizing a random matrix given by~\cite{zhen},
	\begin{equation}\label{111}
		\epsilon={\rm 10\%noise}= 0.1\sqrt{2}(rand(numel(t),1)-0.5)(1 + i)
	\end{equation}
In MATLAB program, $numel(t)$ represents the number $N$ of discretized  values of $t$ (refer Appendix) and $rand(numel(t),1)$ returns a $N\times1$ matrix of uniformly distributed random numbers on the open interval $(0,1)$. The term $(1+i)$ in (\ref{111}) signifies that $\psi(x=0,t)$ is perturbed by complex broadband perturbation.
	The results of numerical simulation of the intensity profiles of the dipole soliton (\ref{1.12}), elliptic solitary wave (\ref{1.20}) having the dipole structure within a period for $m=0.5$ and having the quadrupole structure within a period for $m=0.6$ respectively, elliptic solitary wave (\ref{1.23}) having the dipole structure within a period and trigonometric solitary wave (\ref{1.25}) exhibiting the dipole structure within a period, by adding $10\%$ white noise are shown in Figs.2(d), 3(d), 4(d), 5(d), 6(d) respectively. The results reveal that the elliptic solitary wave (\ref{1.20}) having the dipole structure within a period for $m=0.5$ of the Jacobi elliptic function, has very weak instability in its propagation under $10\%$ ((Fig.3(d)) white noise but the quadrupole structure within a period for $m = 0.6$ (Fig.4(d)) shows stable evolution for the values of the parameters considered herein. But the other dipole solitary waves (\ref{1.23}) and (\ref{1.25}) propagate appreciably in a stable manner for the chosen parameters under finite initial white noise perturbations. It is also important to test the robustness of the solutions by violation of the parametric conditions. This is especially essential for experimental implementations where the values of the parameters are always approximate and may also fluctuate \cite{triki}. Here, the violations of the parametric conditions are done in two ways: (i) by adding $0.003$ to all the parameters $a_i, i = 1,2\cdots 7$ (ii) by adding an overall factor $0.001$ to the right side of the parametric conditions 9(a) - (d), (10), (13), (15), (17). The resulting numerical simulations are shown in Fig.2(e)-(f), Fig.3(e)-(f) $\&$ Fig.4(e)-(f), Fig.5(e)-(f), Figs.6(e)-(f) for the solitary waves (\ref{1.12}), (\ref{1.20}) for  $m=0.5~\&~m=0.6$, (\ref{1.23}), (\ref{1.25}) respectively. It is evident from the above mentioned figures that the solitary waves are quite robust in both the cases of violations of the parametric conditions. It is to be noted that the nonlinear modes are stable under $10\%$ white noise perturbations and also under violations of the parametric conditions for both positive (Figs. 3, 4, 6 ) and negative (Figs. 1, 2, 5) $k$ values (similar to \cite{kanna}) when the other parameters are kept fixed at the chosen values. The positive and negative $k$ values signify that the direction of the wave vector $\vec{k}$ (in one dimension it is a scalar) for forward propagating wave is opposite to that of the backward propagating wave.
	
		\section {Conclusion}\label{con}
		In this article, we have studied the cubic NLH  system in the presence of TOD, FOD and also non-Kerr nonlinearity like the SS and the SFS. This system models ultrashort nonparaxial pulse propagation with Kerr and non-Kerr nonlinearities, higher order dispersions, and spatial dispersion which becomes important when SVEA fails. We have demonstrated that this system features dipole soliton and also elliptic waves having the dipole/quadrupole structure within a period. The novelty of the propagation dynamics arising from incorporating the higher order dispersion and non-Kerr nonlinearity lies in the appearance of dipole structure within a period of these elliptic waves. Most interestingly, for one of the elliptic waves, there is a transition from the dipole to the quadrupole structure when the value of the modulus parameter $m$ of the Jacobi elliptic function $m > 0.5$. Also an arbitrary amplitude sinusoidal periodic wave is obtained which exhibits dipole structure within a period. So far as we are aware of, this interesting characteristic feature of the propagation dynamics of nonparaxial periodic waves has not been unveiled till now. These solitary wave solutions exist when the model parameters satisfy certain parametric conditions. These conditions exhibit a subtle balance between dispersion, Kerr, and non-Kerr nonlinearities, which is essential to control the dynamics of these solitary waves. To determine the physical significance of the obtained solutions, the effect of the nonparaxial parameter on the amplitude, pulse-width and speed of solitary waves are investigated. It has been observed that the speed of the solitary waves can be reduced by tuning the nonparaxial parameter. To test the stability of the solitary waves, numerical simulations have been performed by adding $10\%$ white noise to the initial stationary mode. It has been found that the elliptic wave having the dipole structure (depending on the value of the modulus parameter of Jacobi elliptic function) within a period has very weak instability in its propagation pattern under these noises for the chosen values of the parameters while the corresponding quadrupole structure propagate stably. All the other dipole solitary waves has stable propagation under these white noise perturbations. Furthermore, robustness of the solutions are examined by violation of the parametric conditions in two different ways namely, by adding $.003$ to all the parameters $a_i, i = 1,2 \cdots 7$  and also by adding an overall factor $.001$ to the parametric conditions. The solitary waves are found to be quite robust. The applications of dipole and quadrupole solitary wave solutions presented here may be found in nonparaxial optical contexts involving sub picosecond or femtosecond pulses, such as in ultrahigh-bit-rate optical communication systems \cite{krug72}. 
	The results presented in this article constitute a step forward to the exploration of dipole, multipole, kink, anti-kink etc. solitary waves for inhomogeneous higher order NLH system as is done in \cite{yang} for HNLSE . Also, it would be interesting to construct these types of solitary wave soluions in the presence of quintic, septic Kerr and non-Kerr nonlinearity as obtained in \cite{sarma} and \cite{azz1,triki} respectively for HNLSE . This is important in order to keep up with current developments in high-repetition-rate (beyond ultrashort) fiber-based pulse sources \cite{had}. Some of these issues are  currently being investigated, and we hope to report these in the near future.\\ 
{\bf Data Availability}\\	
	Data sharing is not applicable to this article as no new data were created or analyzed in this study.\\ 
{\bf Conflict of Interest}\\
The authors have no conflicts to disclose.	
~~\\
	
{\bf Appendix :}\\
For checking the dynamical evolution of the derived solutions, the finite difference method \cite{tref} is used with
 $\delta t=0.1$, $\delta x=10^{-4}$. 
and the boundary condition is given by:
$$\psi(x,t)=0,~\text{at the boundary.}$$
A one-dimensional uniform spatial grid is taken consisting of $N$ points labeled by $t_j = -L+(j-1)\delta t$, $j = 1, 2, .. N$, where $\delta t = \frac{2L}{N-1}$ is the lattice spacing and $L$ is the half-width. The left and right boundary points are located at $j = 1$ and $j = N$ respectively. So, the boundary conditions at the edges of the spatial grid are 
$$\psi(x,\pm L)=0.$$
The derivatives are discretized as \cite{tref}
\begin{gather*}
	\psi_x(x,t)|_{(x_i,t_j)}=\frac{\psi_{i+1,j}-\psi_{i-1,j}}{2\delta x},\\
	\psi_{xx}(x,t)|_{(x_i,t_j)}=\frac{\psi_{i+1,j}-2\psi_{i,j}+\psi_{i-1,j}}{\delta x^2},\\
	\psi_t(x,t)|_{(x_i,t_j)}=\frac{\psi_{i,j+1}-\psi_{i,j-1}}{2\delta t},\\
	\psi_{tt}(x,t)|_{(x_i,t_j)}=\frac{\psi_{i,j+1}-2\psi_{i,j}+\psi_{i,j-1}}{\delta t^2},\\
	\psi_{ttt}(x,t)|_{(x_i,t_j)}=\frac{\psi_{i,j+2}-2\psi_{i,j+1}+2\psi_{i,j-1}-\psi_{i,j-2}}{2\delta t^3},\\
	\psi_{tttt}(x,t)|_{(x_i,t_j)}=\frac{\psi_{i,j+2}-4\psi_{i,j+1}+6\psi_{i,j}-4\psi_{i,j-1}+\psi_{i,j-2}}{\delta t^4},\\
\end{gather*}
where $\psi(x_i,t_j)=\psi_{i,j}.$

\end{document}